\documentclass[aps,prb,twocolumn,superscriptaddress,showpacs]{revtex4-1}
\usepackage{amsmath,amssymb,amsfonts,amsthm}
\usepackage{graphicx}
\usepackage{bm}
\usepackage{bbm}
\usepackage{color}
\usepackage{dcolumn}   
\usepackage{epstopdf}
\usepackage[caption=false]{subfig}
\usepackage{color}
\usepackage[colorlinks=true,linkcolor=blue,urlcolor=blue,citecolor=blue]{hyperref}

\begin{document}

 \newcommand{\breite}{1.0} 

\newtheorem{prop}{Proposition}
\newtheorem{cor}{Corollary} 

\newcommand{\be}{\begin{equation}}
\newcommand{\ee}{\end{equation}}

\newcommand{\bea}{\begin{eqnarray}}
\newcommand{\eea}{\end{eqnarray}}
\newcommand{\lt}{<}
\newcommand{\gt}{>}

\newcommand{\Reals}{\mathbb{R}}     
\newcommand{\Com}{\mathbb{C}}       
\newcommand{\Nat}{\mathbb{N}}       

\newcommand{\id}{\mathbboldsymbol{1}}    

\newcommand{\Real}{\mathop{\mathrm{Re}}}
\newcommand{\Imag}{\mathop{\mathrm{Im}}}

\def\O{\mbox{$\mathcal{O}$}}   
\def\F{\mathcal{F}}			
\def\sgn{\text{sgn}}

\newcommand{\deo}{\ensuremath{\Delta_0}}
\newcommand{\dea}{\ensuremath{\Delta}}
\newcommand{\ak}{\ensuremath{a_k}}
\newcommand{\ad}{\ensuremath{a^{\dagger}_{-k}}}
\newcommand{\sx}{\ensuremath{\sigma_x}}
\newcommand{\sz}{\ensuremath{\sigma_z}}
\newcommand{\spl}{\ensuremath{\sigma_{+}}}
\newcommand{\smi}{\ensuremath{\sigma_{-}}}
\newcommand{\alk}{\ensuremath{\alpha_{k}}}
\newcommand{\bk}{\ensuremath{\beta_{k}}}
\newcommand{\ok}{\ensuremath{\omega_{k}}}
\newcommand{\vd}{\ensuremath{V^{\dagger}_1}}
\newcommand{\vi}{\ensuremath{V_1}}
\newcommand{\vo}{\ensuremath{V_o}}
\newcommand{\zc}{\ensuremath{\frac{E_z}{E}}}
\newcommand{\xc}{\ensuremath{\frac{\Delta}{E}}}
\newcommand{\xd}{\ensuremath{X^{\dagger}}}
\newcommand{\aok}{\ensuremath{\frac{\alk}{\ok}}}
\newcommand{\tpw}{\ensuremath{e^{i \ok s }}}
\newcommand{\tpe}{\ensuremath{e^{2iE s }}}
\newcommand{\tmw}{\ensuremath{e^{-i \ok s }}}
\newcommand{\tme}{\ensuremath{e^{-2iE s }}}
\newcommand{\epls}{\ensuremath{e^{F(s)}}}
\newcommand{\emis}{\ensuremath{e^{-F(s)}}}
\newcommand{\epl}{\ensuremath{e^{F(0)}}}
\newcommand{\emi}{\ensuremath{e^{F(0)}}}

\newcommand{\lr}[1]{\left( #1 \right)}
\newcommand{\lrs}[1]{\left( #1 \right)^2}
\newcommand{\lrb}[1]{\left< #1\right>}
\newcommand{\nbt}{\ensuremath{\lr{ \lr{n_k + 1} \tmw + n_k \tpw  }}}

\newcommand{\om}{\ensuremath{\omega}}
\newcommand{\dw}{\ensuremath{\Delta_0}}
\newcommand{\wbp}{\ensuremath{\omega_0}}
\newcommand{\dv}{\ensuremath{\Delta_0}}
\newcommand{\vbp}{\ensuremath{\nu_0}}
\newcommand{\vplus}{\ensuremath{\nu_{+}}}
\newcommand{\vminus}{\ensuremath{\nu_{-}}}
\newcommand{\wplus}{\ensuremath{\omega_{+}}}
\newcommand{\wminus}{\ensuremath{\omega_{-}}}
\newcommand{\uv}[1]{\ensuremath{\mathbf{\hat{#1}}}} 
\newcommand{\abs}[1]{\left| #1 \right|} 
\newcommand{\avg}[1]{\left< #1 \right>} 
\let\underdot=\d 
\renewcommand{\d}[2]{\frac{d #1}{d #2}} 
\newcommand{\dd}[2]{\frac{d^2 #1}{d #2^2}} 
\newcommand{\pd}[2]{\frac{\partial #1}{\partial #2}} 
\newcommand{\pdd}[2]{\frac{\partial^2 #1}{\partial #2^2}} 
\newcommand{\pdc}[3]{\left( \frac{\partial #1}{\partial #2}
 \right)_{#3}} 
\newcommand{\ket}[1]{\left| #1 \right>} 
\newcommand{\bra}[1]{\left< #1 \right|} 
\newcommand{\braket}[2]{\left< #1 \vphantom{#2} \right|
 \left. #2 \vphantom{#1} \right>} 
\newcommand{\matrixel}[3]{\left< #1 \vphantom{#2#3} \right|
 #2 \left| #3 \vphantom{#1#2} \right>} 
\newcommand{\grad}[1]{{\nabla} {#1}} 
\let\divsymb=\div 
\renewcommand{\div}[1]{{\nabla} \cdot \boldsymbol{#1}} 
\newcommand{\curl}[1]{{\nabla} \times \boldsymbol{#1}} 
\newcommand{\laplace}[1]{\nabla^2 \boldsymbol{#1}}
\newcommand{\vs}[1]{\boldsymbol{#1}}
\let\baraccent=\= 

\title{Fast preparation of critical ground states using superluminal fronts}

\author{Kartiek Agarwal}
\email{kagarwal@princeton.edu}
\affiliation{Department of Electrical Engineering, Princeton University, Princeton, New Jersey 08540, USA}
\author{R. N. Bhatt}
\affiliation{Department of Electrical Engineering, Princeton University, Princeton, New Jersey 08540, USA}
\author{S. L. Sondhi}
\affiliation{Department of Physics, Princeton University, Princeton, New Jersey 08544, USA}

\date{\today}
\begin{abstract}
We propose a spatio-temporal quench protocol that allows for the fast preparation of ground states of gapless models with Lorentz invariance. Assuming the system initially resides in the ground state of a corresponding massive model, we show that a superluminally-moving `front' that \emph{locally} quenches the mass, leaves behind it (in space) a state \emph{arbitrarily close} to the ground state of the gapless model. Importantly, our protocol takes time $\mathcal{O} \left( L \right)$ to produce the ground state of a system of size $\sim L^d$ ($d$ spatial dimensions), while a fully adiabatic protocol requires time $\sim \mathcal{O} \left( L^2 \right)$ to produce a state with exponential accuracy in $L$. The physics of the dynamical problem can be understood in terms of relativistic rarefaction of excitations generated by the mass front. We provide proof-of-concept by solving the proposed quench exactly for a system of free bosons in arbitrary dimensions, and for free fermions in $d = 1$. We discuss the role of interactions and UV effects on the free-theory idealization, before numerically illustrating the usefulness of the approach via simulations on the quantum Heisenberg spin-chain. 
\end{abstract}
\maketitle

\paragraph*{Introduction.} A central challenge in harnessing the power of artificial quantum matter---for quantum computing and other technological purposes or for theoretical investigation---is that of quantum state preparation. 
While much progress has been made in engineering extremely isolated quantum systems---ultracold atoms in optical lattices~\cite{mazurenko2017cold,cheuk2016observation,schneiderMBL,luschen2016evidence} or traps~\cite{langen2015ultracold,Langen}, nitrogen vacancy centers~\cite{fuchsNV,DuttNV,MaurerNV,ShuskovNV,AgarwalmagneticnoiseNV,choi2017observation}, ion traps~\cite{debnath2016demonstration,zhang2017observation,jurcevic2014observation}, superconducting qubit structures~\cite{barends2016digitized,boixo2013quantum,johnson2011quantum,dwaveannealing}---as these systems grow more complex, it becomes harder to devise equally elaborate tools to manipulate them while maintaining isolation from sources of decoherence. It is thus important to find theoretical answers to how fast and efficiently specific quantum states can be prepared, and the minimum set of knobs required to do so. 

In this regard, adiabatic evolution has served as a basis for many investigations (cf. Ref.~\cite{griffiths1995introduction}). In its simplest form, the idea is to prepare the system in an eigenstate of a Hamiltonian that is easily accessible and subsequently tune the Hamiltonian slowly so that this eigenstate evolves into the target state. The limitation of this approach is speed; to avoid exciting the system in the process, the time taken must be of the order of the inverse-square of the smallest instantaneous spectral gap between the target and excited states, a quantity which diverges in the thermodynamic limit for many systems of interest. 

To achieve faster state preparation, recent work has proposed engineering counter-diabatic drives~\cite{jarzynskitransitionless,glaser1998unitary,sels2016minimizing} that counter the production of excitations during adiabatic evolution, or more radically, introducing `optimum-control' protocols~\cite{van2016optimal,Jiangfeng2016optimalspinqubit,Superadiabaticspinchaintransfer2017,aashishcleark_diabatic} (including `bang-bang' protocols~\cite{speedlimitHegerfeldt,bao2017optimal,yang2017optimizing,bukov2017machine}) that entirely dispense with the adiabatic ansatz. As of now, a transparent theoretical prescription for diabatic protocols exists only for finite-size systems and how these insights may be extended to generic thermodynamically-large systems (although approximate methods exist for thermodynamically large systems with powerful classical descriptions~\cite{van2016optimal,bulkspinoptimalcontrol}) is unclear. 
Another body of work~\cite{zaletel2016preparation,ho2009universal} has proposed spatial quenches wherein a (large) chunk of the system serves as a bath to suck out entropy from the subsystem of interest. 
\begin{center}
\begin{figure}
\includegraphics[width = 3.2in]{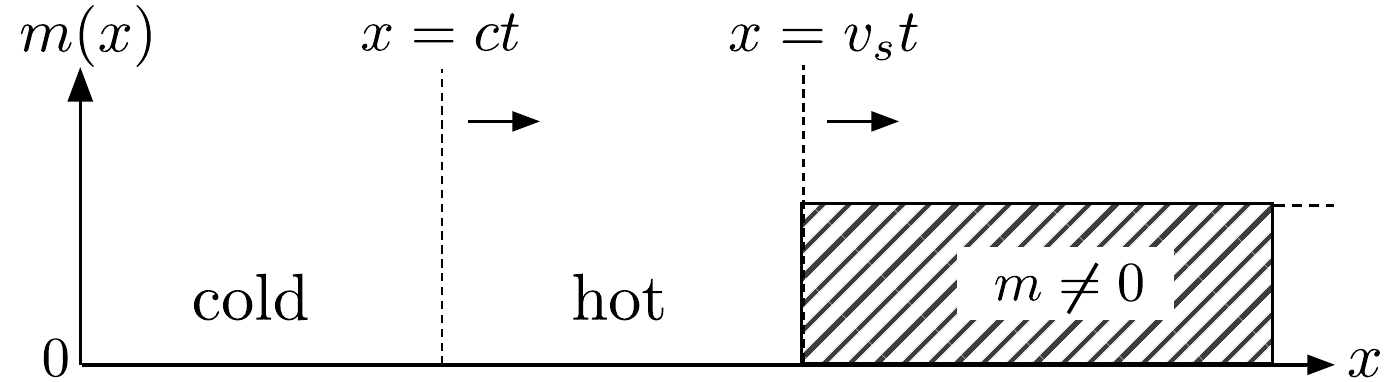}
\caption{The protocol: the local mass is tuned to zero along a front moving at superluminal speed $v_s > c$. As $v_s \rightarrow c^+$, right-moving waves form a shockwave carrying all the energy released in the quench, while the region $ x < c t$, populated only by infinitely red-shifted left-moving waves is left unexcited.}
\label{fig:figone}
\end{figure}
\end{center}

In this letter, we provide a novel example of a diabatic protocol for preparing the ground state of a class of gapless systems---those with emergent Lorentz invariance---starting from the ground state of a corresponding model with an additional mass term that opens up a gap. Such models naturally arise in the low-energy description of various condensed-matter systems, including one-dimensional quantum gases~\cite{Giamarchi} and the Hubbard model at half-filling~\cite{auerbach2012interacting}. We assume that the ground state of the massive model is easier to prepare due to the presence of a gap. Our approach differs from previous approaches inspired by adiabatic evolution in that the method does lead to generation of excitations in the system. Instead our strategy is to invoke the symmetry of the model to `shepherd' excitations in such a way that a thermodynamically large region is left entirely unexcited. 

Specifically, we consider performing a spatio-temporal quench~\cite{AgarwalChiral,Agarwalquantumheatwaves} wherein the \emph{local} mass/gap is tuned to zero---abruptly, or on some time-scale $\tau$---along a superluminal trajectory $x = v_s t$, and uniformly in the other spatial coordinates, as illustrated in Fig.~\ref{fig:figone}. Here the speed $v_s > c$, $c$ being the speed of ``light" in the emergent critical model describing the excitation spectrum $\omega_k = c k$. The quench front then serves as a source of excitations that emanate from the point $x = v_s t$, and travel onwards in all directions. Due to the motion of the front, right-moving excitations get blue-shifted and are populated at higher energies, while left-movers get red-shifted and carry less energy. As the quench speed $v_s$ approaches the speed of sound (from above), the associated Doppler factor diverges and the left-moving modes are left entirely unpopulated. 

In the one-dimensional case, and for non-interacting models, this chirality in excitations has a huge consequence: the region $x < c t$ is only populated by left-moving excitations (right-movers move past into the region between $ x > c t$ and $x < v_s t$) and is left cold. These notions also apply in higher dimensions, albeit with some modifications.  We provide proof-of-concept via an exact solution for the quench problem for free relativistic bosons (in arbitrary dimensions) and fermions (in one dimension) for the locally instantaneous case ($\tau = 0$, finite $v_s$). Our protocol takes time $\sim O(L/c) $ to produce a state that is arbitrarily close to the true ground state of the massless theory. We show that a spatially-uniform, adiabatic protocol ($v_s = \infty$, finite $\tau$) by contrast produces a state that is exponentially close to the ground state in time $\sim \mathcal{O} \left( L^2 \right)$, parametrically slower than our proposed protocol. 

Next, we describe how our results apply to a general setting with interactions, band-curvature and ultraviolet effects. In this regard, we note that the total energy produced in the quench protocol for $\tau = 0$ is, in fact, independent of $v_s$. This implies that cooling occurs only due to spatial segregation of hot and cold regions and may be spoiled by interactions that favor homogenization. We argue that introducing a finite $\tau$ (that does not scale as $L$) solves this issue: a new time-scale $\tau' = \gamma_s \tau$ emerges and controls the adiabaticity of the process. This time-scale diverges in the limit $v_s \rightarrow c$ as the Lorentz dilation factor $\gamma_s = 1/\sqrt{1-1/v^2_s}$ diverges. Thus, the superluminal front enhances the time-scale $\tau$ that introduces adiabaticity. We numerically validate the effectiveness of our protocol via simulations of an anti-ferromagnetic Heisenberg spin chain with a gap induced by local alternating magnetic fields, as well as a classical model of phonons.  

At the time of writing, we became aware of a similar proposal~\cite{Dziarmagainhomogeneous,Dziarmagazgreaterthanone} in the Kibble-Zurek literature wherein a critical velocity for front propagation was proposed using scaling arguments. Our work demonstrates why this critical velocity is exactly $c$ and the importance of relativistic effects in engendering \emph{perfect cooling}. Moreover, Ref.~\cite{Dziarmagainhomogeneous} considers a trans-critical protocol (transforming one gapped state into another, passing through a critical point) as opposed to our work which focusses on the creation of the critical ground state itself. A description of scaling properties of correlations in such inhomogeneous protocols, in the spirit of Ref.~\cite{chandrankibble}, will be discussed in forthcoming work.        

\paragraph*{Model.}

We study the following class of quench models described by the Lagrangian density $\mathcal{L}_b$:
\begin{align}
\mathcal{L}_b &= \partial_\mu \phi \cdot \partial^\mu \phi - m^2 \phi^2 f \left[  \left( x - v_s t \right) / \left( v_s \tau \right) \right],
\label{eq:defmodelmain}
\end{align}
with $ f (x) = \frac{1}{2} \left[ 1 + \text{tanh}( x) \right]$. We set $c \equiv 1 $; $\partial_\mu \equiv (\partial_t , \boldsymbol{\nabla} )$ and $\partial^\mu \equiv (\partial_t, - \boldsymbol{\nabla})$. The function $f$ is such that at time $t = -\infty$ the local mass is finite everywhere, and given by $m$, while at $t = \infty$, the mass vanishes. The fields satisfy the usual commutation relations for relativistic fields~\cite{peskin1995introduction}, and live in $d$ dimensions in a finite box of linear dimension $L$. We assume that the system initially resides in the ground state of the massive model.  

The massless Lagrangian arises naturally in the low-energy limit of a range of one-dimensional systems, including the Heisenberg model~\cite{sirker2012luttinger}, interacting Bose and Fermi gases~\cite{Giamarchi}, or in two-dimensions, for the spin-wave excitations in the Hubbard model at half-filling~\cite{auerbach2012interacting}, etc.. A local gap in spin models may be opened by applying local magnetic fields or dimerization, for instance. 

\paragraph*{Solution for $\tau = 0$, finite $v_s$.} Here, $f(x) = \Theta(x)$, and the solution proceeds as follows. The field operator at all times $t < x / v_s$ can be written in terms of a mode expansion $\phi (x, t < x/v_s) = \sum_n \left[ b_n v_n (x,t) + b^\dagger_n v^*_n (x,t) \right]$, where $v_n$ are solutions to the massive Klein-Gordon equations, 
and $b_n$ are bosonic operators associated with these modes. We work in the Heisenberg picture, fixing the initial state to $\ket{0}$ defined by the condition $b_n \ket{0} = 0 \forall n$. It is important that $v_s > c$; this ensures that no perturbations (that travel at finite speed $c$) due to the quench at $x = v_s t$ affect the space-time region $t < x/v_s$ and the mode expansion above remains valid. 

For times $t > x / v_s$, the evolution of the field operator is described in terms of the massless solutions $u_n$: $\phi(x, t> x/ v_s) = \sum_n \left[ b_n \gamma_n (x,t) + b^\dagger_n \gamma^*_n (x,t) \right]$, where $\gamma_n = \sum_{m} \left[ \alpha_{n,m} u_m + \beta_{n,m} u^*_m \right]$, is determined by matching boundary conditions, that is, $D \gamma_n (x, t = x/v_s) = D v_n (x, t = x/v_s)$ with $D \in \{ 1, \partial_t, \partial_x\}$. The task of finding the `Bogoliubov' coefficients $\alpha_{n,m}$ and $\beta_{n,m}$ is central to the solution of the problem, and is described in detail in the supplemental material (SM). All correlations of the field operators subsequent to the quench can be evaluated using the above expansion, and Wick's theorem applied on the state $\ket{0}$. 
\begin{figure*}
\includegraphics[width = 7in]{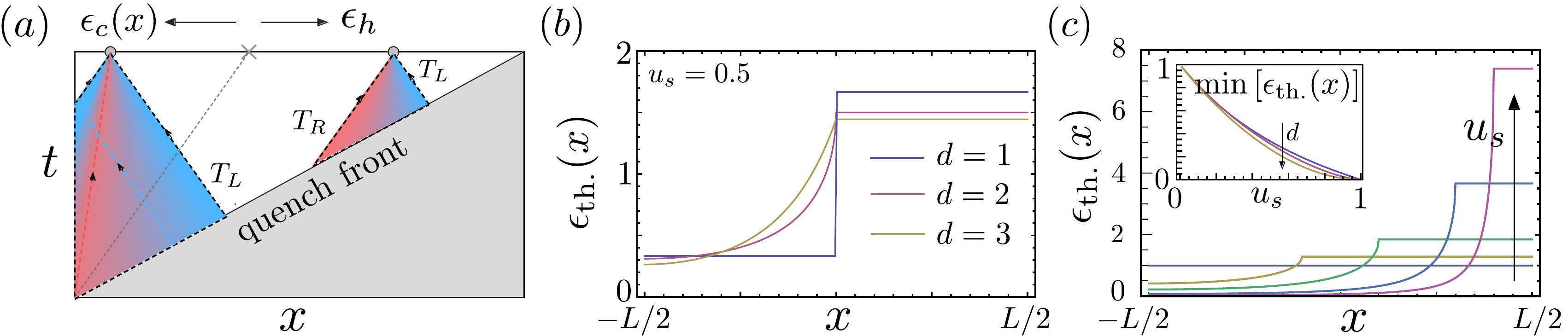}
\caption{(a) The energy density at (x,t) is determined by tracing the excitations impending on it to the quench front. At $t_q$, the end of the quench, $\epsilon_{\text{th.}} (x) $ is uniform in the region $x > c t_q$. (b) $\epsilon_{\text{th.}} (x) $ for $u_s = 0.5$ for various $d$. For $d>1$, the curve is smooth for $x \le c t_q = 0$. (c) $\epsilon_{\text{th.}} (x)$ in $d = 2$, for $u_s = \{0, 0.4, 0.6, 0.8, 0.9 \}$; inset is $min_{x} \epsilon_{\text{th.}} (x) $ as a function of $u_s$ for $d = \{1,2,3\}$. The 'hot' region grows slimmer and the transition into the `cold' region grows sharper as $u_s \rightarrow 1$.}
\label{fig:dimfig}
\end{figure*}
\paragraph*{Chiral emanation.}The energy density after the quench can be evaluated as $\epsilon(x,t) = \sum_n \abs{\nabla \gamma_n}^2 + \abs{\partial_t \gamma_n}^2 \approx \sum_{\vs{k}} \omega_{k} N_{\vs{k}} \abs{u_{\vs{k}} (x,t)}^2$. The last approximate equality is valid in the infinite size limit, neglecting time-dependent terms involving products of wave-functions with two different momenta or terms of the form $u_{k} \cdot u_{-k} \sim e^{- 2i \omega_k t}$ which rapidly dephase. The population of modes is dependent on the direction $\hat{\vs{k}}$ of the mode with momentum $\vs{k}$:

\be
N_{\vs{k}} = \frac{\left( \Omega_{\eta(\theta) k} - \omega_{\eta(\theta) k} \right)^2}{4  \Omega_{\eta(\theta) k} \omega_{\eta(\theta) k}}  \overset{\eta(\theta) \omega_{k} \ll m }{\rightarrow} \frac{m}{4 \eta(\theta) \omega_{k}}
\label{eq:supersonicpop}
\ee 

where $\Omega_k \equiv \sqrt{k^2 + m^2}$, $\omega_k \equiv k$ and $\eta(\theta) = \gamma_s ( 1 - u_s \cos \theta)$ where $u_s = 1/v_s < 1$. $\theta$ ranges from $0$ for right-movers to $\pi$ for left-movers. For the uniform quench ($v_s = \infty$) we note that $\eta (\theta) = 1$. For $v_s \rightarrow c^+$, $\eta(0) \equiv 1/\eta_0 \rightarrow 0$, while $\eta(\pi) = \eta_0 \rightarrow \infty$. This implies that the energy $N_k \omega_k$ carried by left-moving waves goes to zero in this limit. In higher dimensions, most of the emission occurs in directions perpendicular to the motion of the front. Cooling in higher dimensions is based on the fact $\eta ( \pi /2) = \gamma_s$ also diverges in this limit $v_s \rightarrow c^+$; this Doppler shift of orthogonally emitted radiation is a purely relativistic effect.  

\paragraph*{Energy density after quench.} To calculate the full space- and time-dependence of the energy density in the system, slow time-dependent correlations cannot be neglected. Their effect however can be captured using a simple physical picture of `heat waves' as described in Ref.~\cite{Agarwalquantumheatwaves} for $d = 1$, but which we find to be valid generally. In particular, excitations emanate from the quench front, carrying an energy $\omega_k N_{\vs{k}}$ which depends on $\hat{\vs{k}}$. The energy density at any point in space-time is given by the average energy of all excitations emanating from the quench front and ending at this point (numerical verification in SM). Here we focus on the aspect of `cooling' and calculate the energy density at the time the quench ends, $t_q$. First, note that the energy carried by waves emitted in the $\theta$ direction is given by $\epsilon_{\theta} \propto \int^{m/\eta(\theta)} k^{d-1} d k \; \omega_{\vs{k}} N_{\vs{k}} \propto \frac{m}{4} \frac{1}{\eta(\theta)^{d+1}} \frac{1}{L^d_m}$, where $L^d_m = (m/c)^d$ has dimensions of volume. Higher momenta modes yield a parametrically similar contribution. (Note that UV divergences occur for $d \ge 3$ but these can eliminated using finite $\tau$.) In $d = 1$, $\epsilon_{\text{th.}} (x) = \frac{1}{\eta^2_0} \Theta(- x + c t) + \frac{1}{2} \left( \eta^2_0 + \frac{1}{\eta^2_0} \right) \Theta ( x- ct)$; thus, the energy density goes to zero for $x < c t$. Using the picture in Fig.~\ref{fig:dimfig} (a), we find for $d > 1$, 

\begin{align}
&\epsilon_{\text{th.}} \left(x, t_q \right) = \frac{\int_0^{\theta_x} \sin^{d-2} \theta \; \epsilon_\theta + \int_0^{\pi - \theta_x} \sin^{d-2} \theta \; \epsilon_\theta}{\int_0^\pi d \theta \; \sin^{d-2} \theta }, \nonumber \\
&\theta_x = \text{Re} \left[ \text{cos}^{-1} \left( \frac{- (x + L/2) }{u_s L} \right) \right]. 
\label{eq:enerdensitydimensional}
\end{align}
where $\sin^{d-2} \theta$ is the appropriate angular measure in dimension $d > 1$. Some features of the energy density in different dimensions is shown in Fig.~\ref{fig:dimfig}. Importantly, a thermodynamically large portion is always observed to become infinitely cold as $v_s \rightarrow c^+$. 
\begin{figure*}
\includegraphics[width = 7in]{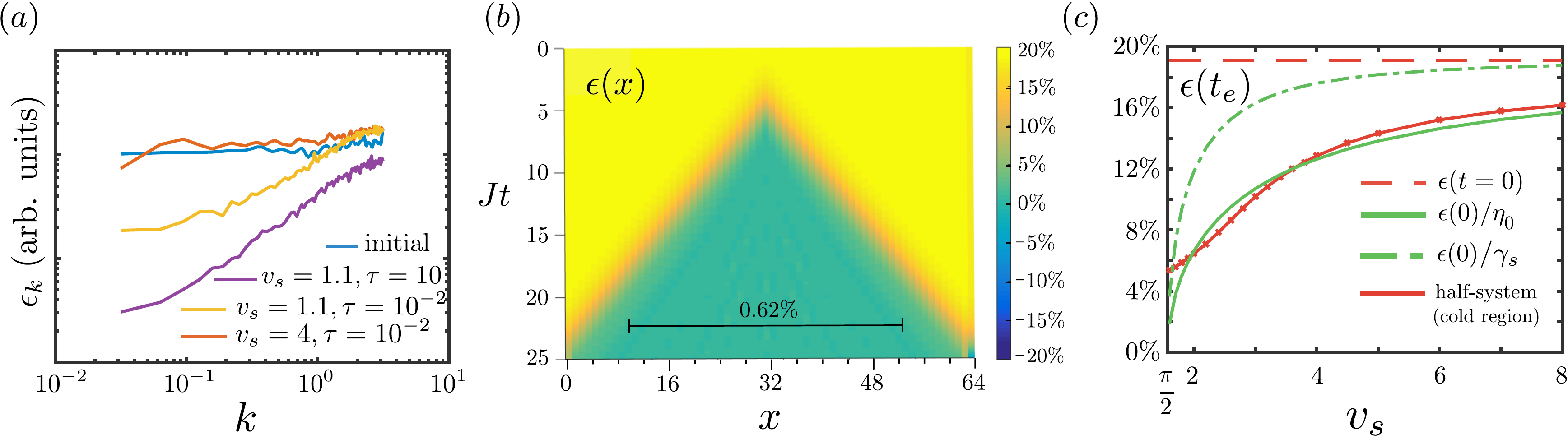}
\caption{(a) Energy per phonon mode at fixed momentum $k$; For large $v_s$ all modes are uniformly excited. As $v_s \rightarrow 1$, low energy `critical' modes are cooled, but UV modes remain unaffected. The population of UV modes can be controlled by increasing $\tau$. (b) t-DMRG simulation of Heisenberg chain for $h_z = 3$, $\tau = 2$, and $v_s/c = 3.2/\pi$. The initial state has energy $\sim 20\%$ of band-width above ground state energy and is readily converted to a state with energy density $\sim 0.62\%$. A linear ramp yields a state that is 3 times hotter in the same time. (c) Quantitative plot of the energy density in the middle half of the chain; $h_z = 3$, $\tau = 0.1$, as a function of $v_s$. The energy of the middle-half of the chain is seen to lie below $\epsilon(0)/\gamma_s$ and follow $\epsilon(0)/\eta_0$ for a large range of $v_s$.}
\label{fig:figreal}
\end{figure*}
\paragraph*{Infinite accuracy.} The above discussion assumed the limit $L \rightarrow \infty$ to find the energy of excitations being emanated in different directions; distinction between `left' and `right' are not valid for modes with momenta $\sim 1/L$. The population of these modes is instead found to scale as $N_k = \frac{m}{4 \gamma_s}$. Importantly, a) this population goes to zero as $v_s \rightarrow c^+$; b) it can be shown that this result is not affected by finite $L$ (a technical discussion and numerical confirmation is presented in SM). Thus, the population of the lowest momentum modes can be tuned arbitrarily close to zero in our protocol.
\paragraph*{Adiabatic Cooling: Solution for finite $\tau$, $v_s = \infty$.} In this case, the quench occurs uniformly in space, but on some time-scale $\tau$. The time-dependent equations of motion can be solved exactly for fixed momenta to find two complete sets of modes $u^{\text{ad.}}_{\vs{k}}$ and $v^{\text{ad.}}_{\vs{k}}$ that behave like the massless and massive modes $u_{\vs{k}}$ and $v_{\vs{k}}$ at $t = \infty$ and $t = -\infty$ respectively---for details of the solution, see Section 3.4 of Ref.~\cite{Birrell} where an analogous problem is solved in the context of a time-dependent metric tensor instead of a time-dependent mass. Thus, one can assume to be initially in the ground state with respect to quanta $b^{\text{ad.}}_{\vs{k}}$ of the modes $v^{\text{ad.}}_{\vs{k}}$. The population of the quanta $a^{\text{ad.}}_{\vs{k}}$ of the modes $u^{\text{ad.}}_{\vs{k}}$, $\avg{a^{\dagger \text{ad.}}_{\vs{k}} a^{\text{ad.}}_{\vs{k}}}$, can be found exactly once $\beta_{\vs{k}}$ in $a^{\text{ad.}}_{\vs{k}} =  \alpha_{\vs{k}} b^{\text{ad.}}_{\vs{k}} - \beta_{\vs{k}} b^{\dagger \text{ad.}}_{(-k_x, \vs{k}_\perp)}$ is determined. We find
\be
N^{\text{ad.}}_{\vs{k}} = \abs{\beta_{\vs{k}}}^2 = \frac{\sinh^2 \left(\frac{\pi}{2} \tau (\Omega_{\vs{k}} - \omega_{\vs{k}}) \right) }{\sinh \left( \pi \tau \omega_{\vs{k}} \right) \sinh \left( \pi \tau \Omega_{\vs{k}} \right)} \overset{k \ll \tau^{-1}}{\longrightarrow} \\ e^{- 2 \pi \omega_k \tau}.
\label{eq:adiabaticpop}
\ee
One can easily check that to obtain an energy density $\epsilon(x) \sim e^{-L}$ in this adiabatic protocol, the time required scales as $\tau_{\text{ad.}} \sim \mathcal{O} ( L^2 )$. Thus, our proposed diabatic protocol which takes time $\sim \mathcal{O} (L)$ is faster and more efficient than the adiabatic protocol.
  
\paragraph*{Bosons vs. fermions.} We note that the fundamental conclusions above are not changed for relativistic theories  with different statistics. We examine this in the context of free relativistic fermions in $d = 1$, governed by the action $\mathcal{L}_f = i \bar{\psi} \partial_\mu \gamma^\mu \psi - m \bar{\psi} \psi f \left[  \left( x - v_s t \right) / \left( v_s \tau \right) \right] $. We again calculate the results for $\tau = 0$, finite $v_s$ by performing an appropriate mode expansion in terms of massive and massless before and after the quench, respectively, and matching the spinor components at the front; details in SM. We find occupation numbers 
\be 
N^F_{\vs{k}} = \frac{\Omega_{\eta(\theta) k} - \omega_{\eta(\theta) k}}{2 \Omega_{\eta (\theta) k}} \overset{\eta ( \theta) k \ll m }{\longrightarrow} \frac{1}{2}. 
\label{eq:chiralpopfermimain}
\ee
Thus, for fermions, we find that excitations are populated up to a `chemical potential' that is Doppler-shifted $\sim m / \eta(\theta)$, as opposed to the bosonic case where the population at low momenta can be captured by a Doppler-shifted effective temperature~\cite{Agarwalquantumheatwaves,CalabreseTransverse2}. 
\paragraph*{Realistic models: Combination of adiabatic and superluminal cooling.} For $\tau = 0$, all cooling occurs due to spatial separation of cold and hot regions---$\int dx \; \epsilon_{\text{th.}} (x, t_q) = 1$, and thus independent of $v_s$. Thus, one may anticipate interactions, which lead to a homogenization of the total energy, spoil the cooling effect. We now provide arguments showing how the introduction of finite $\tau$ resolves this issue. First, we note that the superluminal quench can be analyzed in a Lorentz-boosted frame moving at speed $u_s = 1/v_s < 1$. In this frame, the quench occurs uniformly in space. This analogy is clearly useful for $\tau =0$: one may recover the result of the superluminal quench, Eq.~(\ref{eq:supersonicpop}), using the uniform, adiabatic quench result in Eq.~(\ref{eq:adiabaticpop}) and Doppler shifting the momenta to obtain population of modes in the laboratory frame.  

For large momenta $k \gg 1/L$, we can ignore the breaking of Lorentz symmetry by the walls, and use the above intuition to find the population of modes at finite $\tau$ and $v_s$. In the boosted frame, the mass term transforms as $f \left[ (x - v_s t) / ( v_s \tau ) \right] \rightarrow f \left[ - t' / \tau' \right]$; thus $\tau' = \gamma_s \tau$ emerges as the effective time-scale for the quench in the boosted frame. Doppler-shifting back into the laboratory frame, we find that the population of modes begins to decay exponentially for $\eta ( \theta) \omega_k \ll m, \tau^{-1} /\gamma_s$. Thus, the cut-off in Eq.~(\ref{eq:supersonicpop}) is now set by $m / \gamma_s$ for $\tau^{-1} \approx m$ instead of $m$. This implies that the average energy density $\epsilon_{\text{ss.} + \text{ad.}} = \epsilon_{\text{ss.}} / \gamma^d_s$ goes to zero in the limit $v_s \rightarrow c^+$ for finite $\tau$. This suggests that the protocol can be useful for preparing the ground state of interacting models. 

\paragraph*{UV effects.} The linear dispersion $\omega_k = c k$ is crucial to the Doppler physics we rely on for our diabatic protocol. Beyond a certain energy scale (for instance, set by the lattice), this assumption breaks down and UV modes have a $k$-dependent group velocity $v_g (k) < c$. The effective cooling/heating factor for these modes can be estimated~\cite{Calabrese} as before with $\gamma_s (k) = 1/\sqrt{1 - \left( v_g (k) / v_s \right)^2}$---this is going to deviate minimally from $1$ when $v_g (k)$ is much smaller than $c$. UV modes are thus excited in a non-chiral way. In the event that the UV scale $\Lambda < m / \eta(\theta)$ and $\tau'^{-1} / \eta (\theta)$, we expect the energy density $\epsilon_\theta \sim 1/  \eta( \theta)$, which is yet different from previous cases. In $d = 1$, this predicts a cold region with energy density $m / 4 \eta_0 $ separated from a hot region with energy density $\gamma_s m / 4$; the average energy density is $m / 4 \gamma_s$.    

We next describe simulations of the proposed quench on a classical model of non-interacting phonons to study UV physics, and subsequently study its efficacy on an interacting model---the Heisenberg spin chain. 

\paragraph*{Classical phonons.} We perform simulations for the classical system with Hamlitonian, $H = \frac{1}{2} \sum_i (x_i-x_{i+1})^2 + v^2_i + m^2 x^2_i$ with $m$ being quenched as in Eq.~(\ref{eq:defmodelmain}). Every mode is given an initial energy $\epsilon_k \sim \sqrt{m^2 + 4 \text{sin}^2 (q / 2) }$, akin to the vacuum point energy in an equivalent quantum model. We simulate the dynamics of the system with the superluminally quenched mass. This allows us to verify some of the above expectations for finite $\tau$ and $v_s$; see Fig.~\ref{fig:figreal} (a). 

\paragraph*{Heisenberg chain.} We next perform time-depdendent DMRG simulations (using the iTensor package~\footnote{the ITensor library is available at \href{http://itensor.org}{http://itensor.org}.}) of the anti-ferromagnetic Heisenberg spin chain with a local alternating magnetic field setting the mass. A two-dimensional version of this model applies to the low-energy physics of the half-filled Fermi-Hubbard model, currently of interest in several ultra-cold atom experiments. The Hamiltonian reads $H = J \sum_x \vs{S}_x \cdot \vs{S}_{x+1} + \sum_x (-1)^x h (x,t) $, where the magnetic field $h(x,t)$ is eliminated from the center outward according to the functional form set by $h_z f(x)$ as in Eq.~(\ref{eq:defmodelmain}). In the non-interacting assumption, we expect that the middle region $x \in [ L/2 - c t , L/2 + ct ]$ is illuminated by only cold excitations moving against the front, while the remaining region is illuminated by an admixture of both hot (moving along with the front) and cold excitations---the boundaries between these regions should get blurred due to interactions. In the interest of practicality we focus of the half system in the middle, $x \in [ L/4, 3 L/4]$ and test the claim that the average energy density in this region is below the average energy density $\sim 1/ \gamma_s$ (as expected from above considerations) in the entire system. 
\paragraph*{Summary and conclusions.} In this work, we provided a non-adiabatic method for preparing the ground states of models with Lorentz symmetry. An exact analysis was presented for free relativistic bosons and fermions while analytical arguments and numerical simulations were used to examine the usefulness of our approach for realistic systems with UV effects and interactions. Our protocol should be accessible in experiments in a wide range of setups hosting artificial quantum matter~\cite{Gring2012,zhang2017observation}, particularly ultra-cold atoms~\cite{mazurenko2017cold,cheuk2016observation,brown2016observation,greif2013short}, where an adapted version may serve as an alternate route to preparation of the low-energy state of the Hubbard model at half-filling.    

\paragraph*{Acknowledgements.} We thank Eugene Demler and Emanuele G. Dalla Torre for insightful comments on this work and for previous collaborations. We also thank Ivar Martin, Anatoli Polkovnikov and Alexander M. Polyakov for discussions. We acknowledge support from DOE-BES Grant No. DE-SC0002140 (RNB), DOE Grant No DE-SC/0016244 (SLS) and the U.K. foundation (KA). 


%

\end{document}


\newcommand{\breite}{1.0} 

\newtheorem{prop}{Proposition}
\newtheorem{cor}{Corollary} 

\newcommand{\be}{\begin{equation}}
\newcommand{\ee}{\end{equation}}

\newcommand{\bea}{\begin{eqnarray}}
\newcommand{\eea}{\end{eqnarray}}
\newcommand{\lt}{<}
\newcommand{\gt}{>}

\newcommand{\Reals}{\mathbb{R}}     
\newcommand{\Com}{\mathbb{C}}       
\newcommand{\Nat}{\mathbb{N}}       

\newcommand{\id}{\mathbboldsymbol{1}}    

\newcommand{\Real}{\mathop{\mathrm{Re}}}
\newcommand{\Imag}{\mathop{\mathrm{Im}}}

\def\O{\mbox{$\mathcal{O}$}}   
\def\F{\mathcal{F}}			
\def\sgn{\text{sgn}}

\newcommand{\deo}{\ensuremath{\Delta_0}}
\newcommand{\dea}{\ensuremath{\Delta}}
\newcommand{\ak}{\ensuremath{a_k}}
\newcommand{\ad}{\ensuremath{a^{\dagger}_{-k}}}
\newcommand{\sx}{\ensuremath{\sigma_x}}
\newcommand{\sz}{\ensuremath{\sigma_z}}
\newcommand{\spl}{\ensuremath{\sigma_{+}}}
\newcommand{\smi}{\ensuremath{\sigma_{-}}}
\newcommand{\alk}{\ensuremath{\alpha_{k}}}
\newcommand{\bk}{\ensuremath{\beta_{k}}}
\newcommand{\ok}{\ensuremath{\omega_{k}}}
\newcommand{\vd}{\ensuremath{V^{\dagger}_1}}
\newcommand{\vi}{\ensuremath{V_1}}
\newcommand{\vo}{\ensuremath{V_o}}
\newcommand{\zc}{\ensuremath{\frac{E_z}{E}}}
\newcommand{\xc}{\ensuremath{\frac{\Delta}{E}}}
\newcommand{\xd}{\ensuremath{X^{\dagger}}}
\newcommand{\aok}{\ensuremath{\frac{\alk}{\ok}}}
\newcommand{\tpw}{\ensuremath{e^{i \ok s }}}
\newcommand{\tpe}{\ensuremath{e^{2iE s }}}
\newcommand{\tmw}{\ensuremath{e^{-i \ok s }}}
\newcommand{\tme}{\ensuremath{e^{-2iE s }}}
\newcommand{\epls}{\ensuremath{e^{F(s)}}}
\newcommand{\emis}{\ensuremath{e^{-F(s)}}}
\newcommand{\epl}{\ensuremath{e^{F(0)}}}
\newcommand{\emi}{\ensuremath{e^{F(0)}}}

\newcommand{\lr}[1]{\left( #1 \right)}
\newcommand{\lrs}[1]{\left( #1 \right)^2}
\newcommand{\lrb}[1]{\left< #1\right>}
\newcommand{\nbt}{\ensuremath{\lr{ \lr{n_k + 1} \tmw + n_k \tpw  }}}

\newcommand{\om}{\ensuremath{\omega}}
\newcommand{\dw}{\ensuremath{\Delta_0}}
\newcommand{\wbp}{\ensuremath{\omega_0}}
\newcommand{\dv}{\ensuremath{\Delta_0}}
\newcommand{\vbp}{\ensuremath{\nu_0}}
\newcommand{\vplus}{\ensuremath{\nu_{+}}}
\newcommand{\vminus}{\ensuremath{\nu_{-}}}
\newcommand{\wplus}{\ensuremath{\omega_{+}}}
\newcommand{\wminus}{\ensuremath{\omega_{-}}}
\newcommand{\uv}[1]{\ensuremath{\mathbf{\hat{#1}}}} 
\newcommand{\abs}[1]{\left| #1 \right|} 
\newcommand{\avg}[1]{\left< #1 \right>} 
\let\underdot=\d 
\renewcommand{\d}[2]{\frac{d #1}{d #2}} 
\newcommand{\dd}[2]{\frac{d^2 #1}{d #2^2}} 
\newcommand{\pd}[2]{\frac{\partial #1}{\partial #2}} 
\newcommand{\pdd}[2]{\frac{\partial^2 #1}{\partial #2^2}} 
\newcommand{\pdc}[3]{\left( \frac{\partial #1}{\partial #2}
 \right)_{#3}} 
\newcommand{\ket}[1]{\left| #1 \right>} 
\newcommand{\bra}[1]{\left< #1 \right|} 
\newcommand{\braket}[2]{\left< #1 \vphantom{#2} \right|
 \left. #2 \vphantom{#1} \right>} 
\newcommand{\matrixel}[3]{\left< #1 \vphantom{#2#3} \right|
 #2 \left| #3 \vphantom{#1#2} \right>} 
\newcommand{\grad}[1]{{\nabla} {#1}} 
\let\divsymb=\div 
\renewcommand{\div}[1]{{\nabla} \cdot \boldsymbol{#1}} 
\newcommand{\curl}[1]{{\nabla} \times \boldsymbol{#1}} 
\newcommand{\laplace}[1]{\nabla^2 \boldsymbol{#1}}
\newcommand{\vs}[1]{\boldsymbol{#1}}
\let\baraccent=\= 

\title{Supplemental Material: Fast preparation of critical ground states using superluminal fronts}

\author{Kartiek Agarwal}
\email{kagarwal@princeton.edu}
\affiliation{Department of Electrical Engineering, Princeton University, Princeton, New Jersey 08540, USA}
\author{R. N. Bhatt}
\affiliation{Department of Electrical Engineering, Princeton University, Princeton, New Jersey 08540, USA}
\author{S. L. Sondhi}
\affiliation{Department of Physics, Princeton University, Princeton, New Jersey 08544, USA}
 
\maketitle

\section{Formalism for free bosons for $\tau = 0$} 
\label{sec:freeboson}

The problem is specified by the following action and commutation relations:

\begin{widetext}
\begin{align}
& S = \frac{K}{2} \int dt \int^{L/2}_{-L/2} dx \int d^{d-1} \vs{r}_\perp \big[ (\partial_t \phi )^2 - (\partial_x \phi)^2 - (\boldsymbol{\nabla}_{\perp} \phi )^2 - m^2 \Theta(x - v_s t) \phi^2 \big], \nonumber \\
& [\phi (x,t), \pi (x',t)] = i \delta(x - x'), \; [\phi (x,t), \phi(x', t)] = 0, \; [\pi (x,t), \pi (x', t) ] = 0, \; \text{where} \; \pi \equiv K \partial_t \phi. 
\label{eq:probdef}
\end{align}
\end{widetext}

We set the speed of sound, $c = 1$; the parameter $K$ controls fluctuations and determines the form of correlations. We impose the boundary condition $\phi(x = -L/2, t) = \phi(x = L/2, t) = 0$ at the edges of the system. We assume that the field $\phi$ initially resides in the ground state of the massive theory with mass $m$. Due to the light-cone bound on propagation in the system, all correlations along or prior to the space-like front of the quench are unaffected by the quench process. 

\subsection{Method of solution}
\label{sec:bosonsolution}

\subsubsection{General principle.} The field $\phi(x,t)$ prior to the quench ($t < x / v_s$) is described by a mode-expansion in terms of a complete set of solutions of the massive Klein-Gordon equation, $v_{n} (x,t)$ and $v^*_{n} (x,t)$: 

\be
\phi(x,t) = \frac{1}{\sqrt{K}}  \sum_{n} \left[  b_{n} v_{n} (x,t) + b^\dagger_{n} v^*_{n} (x,t) \right].
\ee

Note that, the modes $v_n$ correspond to positive frequency solutions $\sim e^{- i \omega_n t}$, with $ \omega_n > 0$ of the KG equation, while $v^*_n$ correspond to negative frequency solutions. The coefficients $b_n$ and $b^\dagger_n$ associated with these modes satisfy the bosonic commutation relations, $[b_n, b^\dagger_{m}] = \delta_{n,m}$, $[b^{(\dagger)}_n, b^{(\dagger)}_{m}] = 0$. These define the initial state via the relation $b_n \ket{0} = 0$ for all $n$, that is, the state is a vacuum of all massive excitations.  

After the quench the field operator evolves according to the massless KG equation. Thus, the above mode expansion is not valid for $t > x / v_s$. In order to find correlations for subsequent times, it is useful to find an expansion of the massive modes, $v_n (x,t)$, in terms of a complete basis of massless solutions, $u_m (x,t)$, and $u^*_m (x,t)$, that holds at the quench front, $x = v_s t$. In particular, define the `Bogoliubov' coefficients $\alpha_{n,m}$ and $\beta_{n,m}$ (this terminology is standard in the literature on quantum fields in curved space-time, see Ref.~\cite{Birrell}) such that 

\be 
D v_n \big|_{x = v_s t} = \sum_m \left[ \alpha^*_{n,m} D u_m  - \beta_{n,m} D u^*_m \right] \big|_{x = v_s t}
\label{eq:fieldtransform}
\ee

where $D = 1, \partial_x,$ or $\partial_t$. Then the evolution of the field operator for times $t > x/v_s$ can be described by the expansion 

\begin{align}
\phi(x, t > x/v_s) = \frac{1}{\sqrt{K}} \sum_n \left[ \gamma_{n} (x,t) b_n + \gamma^*_n (x,t) b_n \right], \nonumber \\
 \text{where} \; \gamma_n (x,t) = \sum_m \left[ \alpha^*_{n,m} u_m (x,t) + \beta_{n,m} u^*_m (x,t) \right].
 \end{align}
 
The correlations of the field operators at space-time points $(x,t)$ with $ t > x/v_s$ can be easily evaluated using this expansion and applying Wick's theorem with the state $\ket{0}$. 

\subsubsection{KG inner product and normalization of modes.} 

To normalize the modes and determine the coefficients $\alpha_{n,m}$ and $\beta_{n,m}$, we use the Klein-Gordon inner product~\cite{Birrell,unruheffectandapplications} which is defined in a covariant form as follows: for two solutions $\phi_a$ and $\phi_b$ of the KG equations, the inner product is 

\begin{align}
(\phi_a,\phi_b) &= -i \int d^d \vs{r} \sqrt{g} n_\mu J^{\mu}_{(a,b)} (\vs{r}), \nonumber \\
\text{where}, J^\mu_{(a,b)} &= (\phi_a \nabla^\mu \phi^*_b - \phi^*_b \nabla^\mu \phi_a). 
\label{eq:kginner}
\end{align} 
 
Here the integral is over all space, $g$ is the determinant of the induced metric on space-like coordinates and $\sqrt{g} d^d \vs{r}$ is the covariant volume element, $n^\mu$ is a future-directed time-like unit vector normal to the space-like hypersurface, and $J^\mu_{(a, b)}$ is called the Klein-Gordon current. If $\phi_a$ and $\phi_b$ satisfy the \emph{same} Klein-Gordon equation (massive or massless), then it is easy to check that $\nabla_\mu J^\mu_{(a,b)} = 0$. Thus, the Klein-Gordon current is associated with a conservation law: the integral of the charge, $n_\mu J^\mu$ over all space, is constant over time. We note additionally: 

(a) If the modes $v_n$ form a complete set of modes according to the KG inner product, that is, $(v_n, v_m) = \delta_{n,m}$,  $(v_n, v^*_m) = 0$, and the mode operators $b_n$ satisfy $[b_n,b^\dagger_m] = \delta_{n,m}$, $[b^{(\dagger)}_n,b^{(\dagger)}_m] = 0$, then the field operators (and its conjugate) satisfy the correct commutation relations as described in Eq.~(\ref{eq:probdef}). A proof of this is discussed in Ref.~\cite{}


(b) From its formulation in Eq.~(\ref{eq:kginner}), it is explicit that the KG inner product is invariant under transformation into a coordinate system which admits a separation between time-like and space-like coordinates, that is, the metric is of the form $ds^2 = [N(x,t)]^2 dt^2 - g_{ab} (x,t) dx^a dx^b$. Thus, the normalization relations $(u_n, u_m) = \delta_{n,m}$, $(v_n, v_m) = \delta_{n,m}$, etc. are invariant under such coordinate transformations. 

(c) The above two properties imply that under a Lorentz transformation of the coordinates (without any change in the operators $b_n$, $b^{(\dagger)}_n$), the field operators continue to satisfy the commutation relations in Eq.~(\ref{eq:probdef}) in the \emph{transformed} coordinates as appropriate for a relativistic field. 

(d) Note that the KG inner product additionally satisfies $(\phi_a, \phi_b) = - (\phi^*_a, \phi^*_b)$. Thus, while particle modes satisfy $(u_n, u_m) = \delta_{n,m}$, anti-particle modes satisfy $(u^*_n, u^*_m) = - \delta_{n,m}$. 

\subsubsection{Determination of $\alpha_{n,m}$ and $\beta_{n,m}$.} 

Note that the quench front $x = v_s t$ corresponds to a constant time $t' = 0$ in the Lorentz-boosted coordinate frame $x' = \gamma_s ( x - u_s t )$, $t' = \gamma_s (t - u_s x)$ (with the other coordinates remaining the same) where $\gamma_s = 1/\sqrt{1-u^2_s}$. This corresponds to a boost by velocity $u_s$ is which is the inverse of the super-sonic velocity $v_s$ and is therefore a legitimate Lorentz transformation. 

The KG inner product evaluated at time $t' = 0$ in this frame reads

\be
(\phi_a , \phi_b) = -i \int_{-L/2\gamma_s}^{L/2 \gamma_s} dx' \left[ \phi_a \partial_{t'} \phi^*_b - \phi^*_b \partial_{t'} \phi_a \right] \big|_{x', t' = 0}. 
\label{eq:KGboosted}
\ee

Assuming all parts of Eq.~(\ref{eq:fieldtransform}) hold, one may evaluate $(u_n, v_m)$ and $(u_n, v^*_m)$ (substituting $v_n$ and $\partial_{t'} v_n$ into the definition) to find

\begin{align}
(u_n, v_m)\big|_{t' = 0} &= \sum_l (u_n, \alpha^*_{m,l} u_l - \beta_{m,l} u^*_l) \big|_{t' = 0} = \alpha_{n,m}, \nonumber \\
(u_n, v^*_m)\big|_{t'  =0}  &= \sum_l (u_n, \alpha_{m,l} u^*_l - \beta^*_{m,l} u_l) \big|_{t' = 0} = - \beta_{n,m}, 
\label{eq:alphabeta}
\end{align}

where we used $(u_n, u^*_m)\big|_{t' = 0} = 0$, $(u_n, u_m) \big|_{t' = 0} = \delta_{n,m}$. 

The above suggest that \emph{if} Eqs.~(\ref{eq:fieldtransform}) are simultaneously satisfiable, \emph{then} the coefficients $\alpha_{n,m}$ and $\beta_{n,m}$ must be given by Eq.~(\ref{eq:alphabeta}). In order to confirm that these are indeed the correct solutions, we can substitute these solutions into Eqs.~(\ref{eq:fieldtransform}); then using the commutation relations on the field operators [as in Eq.~(\ref{eq:probdef})] at $t' = 0$ directly confirms the validity of the result. 

We also note the following properties that are satisfied by the Bogoliubov coefficients: 

\begin{align}
(v_n, v_m) &= \sum_{a} \left[ \alpha^*_{a,n} \alpha_{a,m} - \beta_{a,n} \beta^*_{a,m} \right] = \delta_{n,m} \nonumber \\
(v_n, v^*_m) &= \sum_{a} \left[ - \alpha^*_{a,n} \beta_{a,m} + \beta_{a,n} \alpha^*_{a,m} \right] = 0
\label{eq:bogorelations}
\end{align}

 \subsection{Solution of problem}
 
 We now provide explicit details of the solution of the problem posed in Eqs.~(\ref{eq:probdef}). 
 
 \subsubsection{Normalized modes.} 
 
 The massive and massless solutions of the KG-equation satisfying the boundary conditions $\phi(x = \pm L/2 , 0) = 0$ read
 
\begin{align}
u_{\pm, \vs{k}} &= \frac{1}{\sqrt{4 \omega_{\vs{k}} L^{d-1}_\perp L }} e^{i \vs{k}_\perp . \vs{r}_\perp} \left( e^{i k_x x - i \omega_{\vs{k}} t} \pm e^{- i k_x x - i \omega_{\vs{k}} t} \right), \nonumber \\
v_{\pm, \vs{k}} &= \frac{1}{\sqrt{4 \Omega_{\vs{k}} L^{d-1}_\perp L }} e^{i \vs{k}_\perp . \vs{r}_\perp} \left( e^{i k_x x - i \Omega_{\vs{k}} t} \pm e^{- i k_x x - i \Omega_{\vs{k}} t} \right), 
\end{align}

where $\omega_{\vs{k}} = \abs{\vs{k}}$ and $\Omega_{\vs{k}} = \sqrt{\abs{\vs{k}}^2 + m^2}$. The $+ (-) $ modes are associated with momenta $\vs{k} = \left( k_x, \vs{k}_\perp \right) = \left( \frac{ (2 n - 1 )\pi}{ L} \left( \frac{2 n \pi}{L} \right), \frac{2 m \pi}{ L_\perp} \right)$, with $ n \in \mathcal{Z}^+$ and $m \in \mathcal{Z}$. 

\subsubsection{Bogoliubov coefficients.}

The Bogoliubov coefficients can be evaluated by expressing these modes in the Lorentz boosted coordinates $(x',t') = (\gamma_s ( x - u_s t) , \gamma_s ( t - u_s x)$ and evaluating the KG inner product at time $t' = 0$ as per Eq.~(\ref{eq:KGboosted}) and Eqs.~(\ref{eq:alphabeta}). The coefficients read

\begin{widetext}
\begin{align}
\beta^{\epsilon, \epsilon'}_{\vs{k}, \vs{k}'} = - (u_{\epsilon, \vs{k}}, v^*_{\epsilon',  \vs{k}'}) = \frac{1}{4 L \sqrt{\omega_{\vs{k}} \Omega_{\vs{k}'}}} \delta_{\vs{k}_\perp, - \vs{k}'_\perp} &\bigg[ \left( \Omega_R - \omega_R \right) F \left( k_R + k'_R \right) - \epsilon'  \left( \omega_R - \Omega_L \right) F \left( k_R - k'_L \right) \nonumber \\
&- \epsilon  \left( \omega_L - \Omega_R \right) F \left( k'_R - k_L \right) + \epsilon \epsilon'  \left( \Omega_L - \omega_L \right) F \left( -k_L -k'_L \right) \bigg] \nonumber \\
\alpha^{\epsilon, \epsilon'}_{\vs{k}, \vs{k}'} = (u_{\epsilon, \vs{k}}, v_{\epsilon',  \vs{k}'}) =  \frac{1}{4 L \sqrt{\omega_{\vs{k}} \Omega_{\vs{k}'}}} \delta_{\vs{k}_\perp, \vs{k}'_\perp} &\bigg[ \left( \Omega_R + \omega_R \right) F \left( k_R - k'_R \right) + \epsilon'  \left( \omega_R + \Omega_L \right) F \left( k_R + k'_L \right) \nonumber \\
&+ \epsilon  \left( \omega_L + \Omega_R \right) F \left( - k'_R - k_L \right) + \epsilon \epsilon'  \left( \Omega_L + \omega_L \right) F \left( -k_L + k'_L \right) \bigg] \nonumber \\
\label{eq:bogo}
\end{align}
\end{widetext}

where $F(x) = \frac{L}{\gamma_s} \text{sinc} \left( \frac{x L}{2 \gamma_s} \right)$, the Doppler-shifted momenta read 

\begin{align}
k_{R} &= \gamma_s \left(k_x - u_s \omega_{\vs{k}} \right) \in [-u_s \gamma_s \abs{\vs{k}_\perp}, \infty) ,  \nonumber \\
k_{L} &= \gamma_s \left(k_x + u_s \omega_{\vs{k}} \right) \in [u_s \gamma_s \abs{\vs{k}_\perp} , \infty) , \nonumber \\
k'_R &= \gamma_s \left(k_x - u_s \Omega_{\vs{k}} \right) \in [-u_s \gamma_s \sqrt{\abs{\vs{k}_\perp}^2 + m^2}, \infty) , \nonumber \\
k'_L &= \gamma_s \left(k_x + u_s \Omega_{\vs{k}} \right) \in [u_s \gamma_s \sqrt{\abs{\vs{k}_\perp}^2 + m^2}, \infty) ,
\label{eq:Dopplermomenta}
\end{align}

and correspond to frequencies
\begin{align}
\omega_R &=  \sqrt{k^2_R + \abs{\vs{k}_\perp}^2} = \gamma_s \left(\omega_{\vs{k}} - u_s k_x \right), \nonumber \\
\omega_L &=  \sqrt{k^2_L + \abs{\vs{k}_\perp}^2} = \gamma_s \left(\omega_{\vs{k}} + u_s k_x \right), \nonumber \\
\Omega_R &=  \sqrt{k^2_R + \abs{\vs{k}_\perp}^2 + m^2 } = \gamma_s \left(\Omega_{\vs{k}} - u_s k_x \right), \nonumber \\
\Omega_L &=  \sqrt{k^2_L + \abs{\vs{k}_\perp}^2 + m^2 } = \gamma_s \left(\Omega_{\vs{k}} + u_s k_x \right). 
\label{eq:Dopplerfrequencies}
\end{align}

We note these are the usual transformation laws of momenta and frequency under a Lorentz transformation. In what follows, we generally define $\omega (k) = \sqrt{k^2 + \abs{\vs{k}_\perp}^2}$ and $\Omega (k) = \sqrt{k^2 + \abs{\vs{k}_\perp}^2 + m^2}$. Note that, we will also understand the above expressions for $k_R, k_L, k'_R, k'_L, \omega_R, \omega_L, \Omega_R, \Omega_L$ as functions that as input a momentum $\vs{k}$ and spit out the values in Eqs.~(\ref{eq:Dopplermomenta}) and~(\ref{eq:Dopplerfrequencies}). We will also assume that inverse functions can be appropriately defined (as these are monotonic functions). 

\subsubsection{Infinite-size limit and creation of excitations from vacuum.}
\label{sec:infinitsize}

We now examine in detail the creation of excitations due to the quench. This is easiest to appreciate in the infinite-size limit, $L \rightarrow \infty$, but the results discussed below apply generally. In this limit, the function $F(x) \rightarrow 2 \pi \delta (x)$, and it is clear that the decomposition of the massive modes $v_{\pm, \vs{k}}$ in terms of the massless modes occurs by way of momentum matching \emph{in the Lorentz boosted frame}. 

In particular, the mode $v_{\pm, \vs{k}}$ ($u_{\pm, \vs{k}}$), is a linear combinations of a right-moving `particle' with momentum $k'_R$ ($k_R$) and left-moving `particle' with momentum $-k'_L$ ($-k_L$). They also carry a momentum $\vs{k}_\perp$ in the direction orthogonal to the quench front. Concomitantly, the conjugate mode $v^*_{\pm, \vs{k}}$ ($u^*_{\pm, \vs{k}}$) is a linear combination of a right-moving `anti-particle' with momentum $-k'_R$ ($-k_R$) and a left-moving `anti-particle' with momentum $k'_L$ ($k_L$). These modes carry a momentum $-\vs{k}_\perp$ in the orthogonal direction. The Bogoliubov coefficients [Eq.~(\ref{eq:bogo})] are real in the $L \rightarrow \infty$ limit, and reduce to a sum of four momentum-matching delta functions. These delta-functions can be understood by making note of the form of the expansion $v_{\epsilon', \vs{k}'} (t' = 0) = \sum_{\epsilon, \vs{k}} \left[  \alpha^{* \epsilon, \epsilon'}_{\vs{k},\vs{k}'} u_{\epsilon, \vs{k}} -  \beta^{\epsilon, \epsilon'}_{\vs{k},\vs{k}'} u^*_{\epsilon, \vs{k}} \right]  (t' = 0) $, and the momentum carried by the various modes as discussed above. 

The heating up of the system due to the quench occurs due to the conversion of particles into anti-particles and vice-versa. It is thus useful to examine the \emph{massless} anti-particle content ($\propto u^*_{\epsilon, \vs{k}}$) in the \emph{massive} particle mode $v_{\epsilon', k'}$ (or massive anti-particles to massless particles). We find, in particular, (at $t' = 0$)

\begin{align}
v_{\epsilon', \vs{k}'} &= \sqrt{\frac{\omega_{\vs{k}}}{\Omega_{\vs{k}'}}} \left[ u^*_{+, \vs{k}} + u^*_{-, \vs{k}} \right]  \left(\frac{ \omega( k'_R ) - \Omega (k'_R)}{4 \omega (k'_R)} \right) \bigg|_{k = k^{-1}_R (-k'_R)} \nonumber \\
&+ \sqrt{\frac{\omega_{\vs{k}}}{\Omega_{\vs{k}'}}} \left[ u^*_{+, \vs{k}} - u^*_{-, \vs{k}} \right]  \left(\frac{ \omega( k'_R ) - \Omega (k'_R)}{4 \omega (k'_R)} \right) \bigg|_{k = k^{-1}_L (k'_R)} \nonumber \\
& - \sqrt{\frac{\omega_{\vs{k}}}{\Omega_{\vs{k}'}}} \epsilon' \left[ u^*_{+, \vs{k}} + u^*_{-, \vs{k}} \right]  \left(\frac{ \omega( k'_L ) - \Omega (k'_L)}{4 \omega (k'_L)} \right) \bigg|_{k = k^{-1}_R (k'_L)} \nonumber \\
&+ \text{particle content} \propto u_{\pm, \vs{k}}. 
\label{eq:boundarymatch}
\end{align}     

The above is a direct result of the integration over the 4 momentum-conserving delta-functions of the Bogoliubov coefficients; one of these terms, $\propto \delta (-k_L -k'_L)$, does not contribute. Two of these terms are associated with the production of right-movers ($\propto u^*_{+,\vs{k}}  + u^*_{-,\vs{k}}$), and one term is associated with left-movers ($\propto u^*_{+,\vs{k}} - u^*_{-,\vs{k}}$). We focus on the first and second terms---the third term can be shown to be continuously related to the first term, but carries a momentum $k_x > k_0$ while the first carries momentum $k_x < k_0$, with $k_0 = k^{-1}_R ( u_s \gamma_s \sqrt{\abs{\vs{k}_\perp}^2 + m^2})$. 

It is important to note that the amplitude of the first two terms are identical ($\sqrt{\omega_{\vs{k}}}$ is canceled by the same factor in the modes), but the momentum carried by the modes is different. This is not surprising. Left-moving excitations are differently Doppler-shifted to the right-moving ones while the excitation amplitude is only dependent on the absolute value of the momentum in the Lorentz-boosted frame. This results in the differential population and energy carried by left- and right-moving excitations. 

\subsubsection{Doppler-shifted population of left- and right-moving modes.} 


The energy density carried by the field is given by 

\begin{align}
\epsilon (x, t > x/v_s) &= \frac{K}{2} \left\langle \abs{\partial_t \phi}^2 + \abs{\partial_x \phi}^2 + \abs{\boldsymbol{\nabla}_{\perp} \phi}^2 \right\rangle \nonumber \\
&= \frac{1}{2} \sum_{\epsilon', \vs{k}'} \left[ \abs{\partial_t \gamma_{\epsilon', \vs{k}'}}^2 + \abs{\partial_x \gamma_{\epsilon', \vs{k}'}}^2 + \abs{\boldsymbol{\nabla}_{\perp} \gamma_{\epsilon', \vs{k}'}}^2 \right] \nonumber \\
&\approx \sum_{\epsilon', \vs{k}', D} \abs{\sum_{\epsilon, \vs{k}} \beta^{\epsilon,\epsilon'}_{\vs{k}, \vs{k}'} D u^*_{\epsilon, \vs{k}}}^2 + \frac{1}{2} \sum_{\epsilon, \vs{k}, D} \abs{D u_{\epsilon, \vs{k}}}^2 \nonumber \\
&\approx \sum_{\vs{k}} \omega_{\vs{k}} \left[ N_L (\vs{k}) | u_{L, \vs{k}} |^2 + N_R (\vs{k}) | u_{R, \vs{k}} |^2 \right] 
\label{eq:enerdensity}
\end{align}

where $D \in \{\partial_t, \partial_x, \boldsymbol{\nabla}_\perp\}$. To derive the third line in the equation above, we used the first relation in Eq.~(\ref{eq:bogorelations}) and we neglected terms of the form $\sim u^* \cdot u^*$ and $\sim u \cdot u$: these terms are time-dependent and reduce to zero upon averaging at long times. The second term in the third line is the vacuum contribution; it is constant function of space and will be neglected in the fourth line. The internal sum in the first term can be directly replaced with the results of Eq.~(\ref{eq:boundarymatch})---here we define $u_{R / L, \vs{k}} = (u_{+, \vs{k}} \pm u_{-, \vs{k}})/2$. Note that to arrive at the last line, we neglected terms of the form $u^*_{L, \vs{k}_1} u_{R,\vs{k}_2}$ because the momentum carried by these terms is different and these also dephase rapidly. 

Using the invariance of the measures $\sum_{\vs{k}'} \frac{1}{\Omega_{\vs{k}'}}$, $\sum_{\vs{k}} \frac{1}{\omega_{\vs{k}}}$ under Lorentz transformations $\left( \text{for example,}  \sum_{\vs{k}'} \frac{1}{\Omega_{\vs{k}'}} =  \sum_{\vs{k}'_R} \frac{1}{\Omega_R} \right)$, and the results of Eq.~(\ref{eq:boundarymatch}), we arrive at the result for populations $N_{L, \vs{k}}$ and $N_{R, \vs{k}}$ as in the main text. 

We note that the final result for the energy density in Eq.~(\ref{eq:enerdensity}) is valid primarily in the space-time vicinity of the quench front. This is because emanated modes can reflect off the boundaries are change direction---this physics is contained in the slow space and time dependent terms that we have neglected. The easiest way to see this is to note that the momentum carried in $+$ and $-$ modes differs by at least $\pi/L$---this momentum difference becomes important at a time $\sim L/c$, whence a left-moving mode $u_{L, k} \approx (u_{+, k_+} - u_{-, k_-}) /2$ can transform into a right-moving mode because of a flip in the phase between the $+$ and $-$ components. 

The result of Eq.~(\ref{eq:enerdensity}) is interpreted as follows. Since the quench can be viewed as an instantaneous, spatially uniform process in a Lorentz-boosted frame, the massless modes are homogeneously excited by the quench process in this frame. For modes with momenta $\vs{k} \ll m$, the excess vacuum energy of the massive modes, $\sim m$ per mode will be dumped into the new massless excitations. By equipartition, we expect their population to be given by $m / \abs{\vs{k}}$. Now, these modes will appear Doppler-shifted back in the laboratory frame.  

\section{Formalism for Free Fermions for $\tau = 0$}
\label{sec:freefermi}

The problem is specified by the following action and commutation relations:

\begin{widetext}
\begin{align}
& S = \int dt \int^{L/2}_{-L/2} dx \int dx \; \big[ \frac{i}{2} \left( \bar{\psi} \gamma^\alpha \partial_\alpha \psi - \partial_\alpha \bar{\psi} \gamma^\alpha \psi \right) - m \Theta(x - v_s t)  \bar{\psi} \psi \big] \nonumber \\
& \{ \psi^a (x,t), \psi^{\dagger b} (x',t) \} = i \delta(x - x') \delta_{a,b} \;\;\; \text{and}  \;\;\; [\psi^a (x,t), \psi^bb (x', t)] = 0. 
\label{eq:probdeffermi}
\end{align}
\end{widetext}

We again set the speed of sound, $c = 1$. We will work in the Weyl basis wherein $\gamma^0 = \gamma_x$ and $\gamma^1 = -i \sigma_y$, $\sigma_{x,y}$ being Pauli matrices (see section 4.2.4 in Ref.~\cite{fradkin2013field}, and more generally, Ref.~\cite{peskin1995introduction}). $\bar{\psi} = \psi^\dagger \gamma^0$, where $\psi^a (x,t)$ is the field of the Weyl fermion with spinor index $a = 1 , 2$. $\partial_\mu = ( \partial_t, \partial_x)$. Note that we will use superscript for the spinor index while the subscript will be used to distinguish different solutions of the Dirac equation. The quench occurs, as before, locally along a front that propagates towards the right at a fixed, supersonic speed $v_s > 1$. The velocity $u_s \equiv v^{-1}_s < 1$ is defined as before. 

While the Hamiltonian before and after the quench satisfies the usual discrete symmetries associated with free relativistic fermions, imposing a boundary necessitates the breaking of some of these symmetries, see Ref.~\cite{alonso1997diracfermionbox}. We work with `natural' states that forgo parity (P) and charge-conjugate (C) symmetries but keep time-reversal (T) and the combined CPT symmetry. These states also satisfy the condition that the current $c \psi^\dagger \sigma_z \psi$ is zero at the edges of the system. This is effected with the following set of boundary conditions: $\psi_1 (L/2) = \psi_2 (L/2)$, $\psi_1 (-L/2) = -\psi_2 (-L/2)$. One can check for these conditions that $T \psi (x,t) = - \sigma_x \psi^* (x, -t)$ and $CPT \psi (x,t) = - \sigma_z \psi (-x, -t)$ satisfy the same boundary conditions while $P \psi (x,t) = \sigma_x \psi (-x, t)$ and $C \psi (x,t) = \sigma_z  \psi^* (x, t)$ do not.  

\subsection{Method of solution}
\label{sec:fermisolution}

\subsubsection{General principle.}

We follow a similar strategy to the solution of the bosonic case. We work in the Heisenberg picture, and describe the field operator prior to the quench $(t < x / v_s)$ by a mode-expansion in terms of the complete set of solutions of the massive Dirac equation, $i \gamma^\mu \partial_\mu - m) \psi = 0$.  In particular, these are positive-frequency particle (or `electron') modes $v_n (x,t) $, and negative-frequency anti-particle (or `hole') modes $\tilde{v}_n = C v_n = \sigma_z v^*_n$. Thus, 

\be 
\psi (x, t < x/v_s) = \sum_n \left[ f_n v_n (x,t) + \tilde{f}^\dagger_n \tilde{v}_n (x,t) \right]
\ee

The coefficients associated $b_n$ and $\tilde{b}_n$ satisfy the usual fermionic anti-commutation relations: all operators anti-commute besides $\{ f_n , f^\dagger_m \} = \delta_{n,m}$ and $\{ \tilde{f}_n , \tilde{f}^\dagger_m \} = \delta_{n,m}$. The initial state is defined via the relation $f_n \ket{0} = 0$ and $\tilde{f}_n \ket{0} = 0$ for all $n$. Note that this amounts to setting the initial state as being the vacuum of hole-like and particle-like excitations, which is relevant case for a critical system. As before, the above expansion is valid for all times $t < x/v_s$ since this quench occurs on a space-like hypersurface. 

After the quench, the field operator evovles according to the massless KG equation and the mode expansion above is not valid for $t > x / v_s$. To find correlations for subsequent times, we expand the massive modes in terms of the massless modes. We assume the notation $u_n (x,t)$ and $\tilde{u}_n (x,t)$ analogously to before for the massless modes, and define 

\begin{align}
v_n \big|_{x = v_s t} &= \sum_m \left[ \alpha^*_{n,m} u_m + \beta_{n,m} \tilde{u}_m \right] \big|_{x = v_st}, \nonumber \\
\tilde{v}_n \big|_{x = v_s t} &= \sum_m \left[ \alpha_{n,m} \tilde{u}_m + \beta^*_{n,m} u_m \right] \big|_{x = v_st}, 
\label{eq:fermionmatch}
\end{align}

where the second equation follows from the first upon application of the charge-conjugate operation. Then the evolution of the field operator for times $ t > x / v_s$ can be described by the expansion 

\begin{align}
\phi ( x, t > x / v_s) &= \sum_n \left[ \gamma_n (x,t) f_n + \tilde{\gamma}_n (x,t) \tilde{f}^\dagger_n \right], \nonumber \\
\text{where} \; \gamma_n (x,t) &= \sum_m  \left[ \alpha^*_{n,m} u_m (x, t) + \beta_{n,m} \tilde{u}_m (x,t) \right], \nonumber \\
\text{and} \; \tilde{\gamma}_n (x,t) &= \sum_m \left[ \alpha_{n,m} \tilde{u}_m (x, t) + \beta^*_{n,m} u_m (x,t) \right]. 
\end{align}

\subsubsection{Dirac inner product and normalization of modes.}

We use a coordinate-system invariant normalization scheme for the modes which also allows us to determine the Bogoliubov coefficients $\alpha_{n,m}$ and $\beta_{n,m}$. We define the Dirac inner product between two solutions $\psi_a $ and $\psi_b$ as

\begin{align}
(\psi_a, \psi_b) = \int d x \sqrt{g} n_\mu J^\mu_{(a,b)} (x), \nonumber \\
\text{where} \; J^\mu_{(\psi_a, \psi_b)} = \bar{\psi} \gamma^\mu \phi.
\label{eq:diracinner}
\end{align}

Here the integral is over all space, $g$ is the determinant of the induced metric on space-like coordinate, $\sqrt{g} dx $ is the covariant volume element, $n^\mu$ is a future-directed time-like unit vector normal to the space-like hypersurface, and $J^\mu_{(a, b)}$ is the Dirac current. If $\psi_a$ and $\psi_b$ satisfy the \emph{same} Dirac equation (massive or massless), then it is easy to check that $\partial_\mu J^\mu_{(a,b)} = 0$. Thus, the integral over all space of the charge associated with this current, $n_\mu J^\mu$, is constant over time. As before we note: 

(a) If the modes $v_n$ and $\tilde{v}_n$ form a complete set of modes according to the Dirac inner product, that is, $(v_n, v_m) = \delta_{n,m}$,  $(v_n, \tilde{v}_m) = 0$, and the mode operators $f_n$ and $\tilde{f}_n$ satsify the usual fermionic anti-commutation relations, then it can be shown that the field operators (and its conjugate) satisfy the correct commutation relations as described in Eq.~(\ref{eq:probdeffermi}). The proof follows similarly to that given for the bosonic case. 

(b) From its formulation in Eq.~(\ref{eq:diracinner}), it is explicit that the Dirac inner product is invariant under transformation into a coordinate system which admits a separation between time-like and space-like coordinates, that is, the metric is of the form $ds^2 = [N(x,t)]^2 dt^2 - g (x,t) dx^2$. Thus, the normalization relations $(u_n, u_m) = \delta_{n,m}$, $(v_n, v_m) = \delta_{n,m}$, etc. are invariant under such coordinate transformations. 

(c) The above two properties imply that under a Lorentz transformation of the coordinates (without any change in the operators $f_n$, $\tilde{f}_n$), the field operators continue to satisfy the commutation relations in Eq.~(\ref{eq:probdef}) in the \emph{transformed} coordinates, as appropriate for a relativistic field.

(d) The Dirac inner product has the symmetry that $(\psi_a, \psi_b) = (\tilde{\psi}_a, \tilde{\psi}_b)$. Thus, both particle and anti-particle modes follow the same normalization scheme, that is, $(u_n, u_m) = (\tilde{u}_n, \tilde{u}_m ) = \delta_{n,m}$. 


\subsubsection{Determination of $\alpha_{n,m}$ and $\beta_{n,m}$}

The Dirac inner product evaluated at time $t' = 0$ (in the boosted frame coordinates) reads

\be
(\psi_a , \psi_b) = \int_{-L/2\gamma_s}^{L/2 \gamma_s} dx' \; \psi^\dagger_a \psi_b \big|_{t' = 0}.
\label{eq:Diracboosted}
\ee

Assuming Eqs.~(\ref{eq:fermionmatch}) hold at $t' = 0$, one may evaluate $(u_n, v_m)$ and $(u_n, \tilde{v}_m)$ to find

\begin{align}
(u_n, v_m)\big|_{t' = 0} &= \sum_m (u_n, \alpha^*_{n,m} u_m + \beta_{n,m} \tilde{u}_m) \big|_{t' = 0} = \alpha^*_{n,m}, \nonumber \\
(u_n, \tilde{v}_m)\big|_{t'  =0}  &= \sum_m (u_n, \alpha_{n,m} \tilde{u}_m + \beta^*_{n,m} u_m) \big|_{t' = 0} = \beta^*_{n,m}, 
\label{eq:alphabetafermi}
\end{align}

where we used $(u_n, \tilde{u}_m)\big|_{t' = 0} = 0$, $(u_n, u_m) \big|_{t' = 0} = \delta_{n,m}$. The above suggest that \emph{if} Eqs.~(\ref{eq:fermionmatch}) are simultaneously satisfiable, \emph{then} the coefficients $\alpha_{n,m}$ and $\beta_{n,m}$ must be given by Eq.~(\ref{eq:alphabetafermi}). In order to confirm that these are indeed the correct solutions, we can substitute these solutions into Eqs.~(\ref{eq:fermionmatch}); then using the commutation relations on the field operators [as in Eq.~(\ref{eq:probdeffermi})] at $t' = 0$ directly confirms the validity of the result. 

By the methods above, we may also show the inverse expansion at $t' = 0$:

\begin{align}
u_m \big|_{x = v_s t} &= \sum_n \left[ \alpha_{n,m} v_n + \beta_{n,m} \tilde{v}_n \right] \big|_{x = v_st}, \nonumber \\
\tilde{u}_m \big|_{x = v_s t} &= \sum_n \left[ \alpha^*_{n,m} \tilde{v}_n + \beta^*_{n,m} v_n \right] \big|_{x = v_st}.
\label{eq:fermionmatchinverse}
\end{align}

Using these, one can easily prove that these fermionic Bogoliubov coefficients has the following useful property: 

\begin{align}
(u_n, u_m) &= \sum_{a} \left[ \alpha^*_{a,n} \alpha_{a,m} + \beta^*_{a,n} \beta_{a,m} \right] = \delta_{n,m}. \label{eq:bogorelationsfermi}
\end{align}

\subsection{Solution of problem}

We now provide details of the solution of the problem defined in Eq.~(\ref{eq:probdeffermi}). 

\subsubsection{Normalized modes.}

The massive particle modes are defined as

\begin{align}
v_{\pm, k} &= \frac{1}{\sqrt{2L}} \left( v_k \pm i v_{-k} \right) \; \text{for} \; \;  k>0, \nonumber \\
\text{where} \; \; v_{k} &= \begin{pmatrix} \text{cos} \left( \theta_{k}/2 \right) \\ \text{sin} \left( \theta_{k}/2 \right) \end{pmatrix} e^{- i k x + i \Omega_k t}, \nonumber \\
\; \; \tilde{v}_{-k} &= \begin{pmatrix} \text{sin} \left( \theta_{k}/2 \right) \\ \text{cos} \left( \theta_{k}/2 \right) \end{pmatrix} e^{ i k x + i \Omega_k t}, 
\end{align}

while the anti-particle modes are defined as

\begin{align}
\tilde{v}_{\pm, k} &= \frac{1}{\sqrt{2L}} \left( \tilde{v}_k \pm i \tilde{v}_{-k} \right) \; \text{for} \; \;  k>0, \nonumber \\
\text{where} \; \; \tilde{v}_{k} &= \begin{pmatrix} \text{cos} \left( \theta_{k}/2 \right) \\ - \text{sin} \left( \theta_{k}/2 \right) \end{pmatrix} e^{i k x - i \Omega_k t}, \nonumber \\
\; \; v_{-k} &= \begin{pmatrix} \text{sin} \left( \theta_{k}/2 \right) \\ - \text{cos} \left( \theta_{k}/2 \right) \end{pmatrix} e^{- i k x - i \Omega_k t}. 
\end{align}

In the above, $\cos (\theta_k / 2) = \sqrt{\frac{1}{2} + \frac{1}{2} \frac{k}{\Omega_k} }$,  $\sin (\theta_k / 2) = \sqrt{\frac{1}{2} - \frac{1}{2} \frac{k}{\Omega_k} }$, and $\Omega_k = \sqrt{m^2 + k^2}$. The boundary condition is chosen such that CPT symmetry is preserved for the states. A compliant boundary condition corresponds to the choice $\psi_1 (L/2) = \psi_2 (L/2)$ and $\psi_1 (-L/2) = - \psi_2 (-L/2)$. The modes satisfy this boundary condition for $k L = n \pi + \pi /2$, with $n \in [0, 2, 4, ...)$ for modes $v_{+, k}$ and $n \in [1, 3, ...)$ for modes $v_{-, k}$.

An analogous set of massless modes, $u_{\pm, k > 0}$ and $\tilde{u}_{\pm, k > 0}$ can be found by setting the mass to zero in the corresponding formulae for the massive modes. This is enforced by the substitutions $\Omega_k \rightarrow \omega_k = \abs{k}$, $\cos ( \theta_k / 2) \rightarrow 1$, $\sin (\theta_k / 2) \rightarrow 0$. 

It is also useful to note the form of these modes in the Lorentz-boosted frame. The coordinates and momenta are boosted in the usual way, with $k' x - \Omega_{k'} t \rightarrow k'_R x' - \Omega_{k'_R} t'$ and $- k' x - \Omega_{k'} t \rightarrow - k'_L x' - \Omega_{k'_L} t'$. The spinor part is transformed by multiplication with the matrix $\Lambda \equiv -i e^{ \frac{\omega}{2} \cdot \frac{i}{4} [ \gamma^0, \gamma^1 ] } = \begin{pmatrix} 1/\sqrt{\eta} & 0 \\ 0 & \sqrt{\eta} \end{pmatrix}$. Note that $\omega = \text{tanh}^{-1} ( - u_s )$ is the rapidity associated with the Lorentz boosted. 

\subsubsection{Bogoliubov coefficients.}

The Bogoliubov coefficients can be evaluated by expressing these modes in the Lorentz boosted coordinates and evaluating the Dirac inner product at time $t' = 0$ as per Eq.~(\ref{eq:Diracboosted}) and Eqs.~(\ref{eq:alphabetafermi}). The coefficients read

\begin{widetext}
\begin{align}
\beta^{\epsilon, \epsilon'}_{\vs{k}, \vs{k}'} = (u_{\epsilon, \vs{k}}, \tilde{v}_{\epsilon',  \vs{k}'})^* = \frac{1}{2 L} &\bigg[ \frac{\text{cos}\left(\theta_{k'}/2 \right)}{\eta} F \left( k_R + k'_R \right) - i \epsilon' \frac{\text{sin}\left(\theta_{k'}/2 \right)}{\eta} F \left( k_R - k'_L \right) \nonumber \\
&- i \epsilon \; \eta \; \text{sin} \left( \theta_{k'} / 2 \right) F \left( k'_R - k_L \right) - \epsilon \epsilon'  \; \eta \; \text{cos} \left( \theta_{k'} / 2 \right) F \left( -k_L -k'_L \right) \bigg] \nonumber \\
\alpha^{\epsilon, \epsilon'}_{\vs{k}, \vs{k}'} = (u_{\epsilon, \vs{k}}, v_{\epsilon',  \vs{k}'})^* =  \frac{1}{2 L} &\bigg[  \frac{\text{cos}\left(\theta_{k'}/2 \right)}{\eta} F \left( k_R - k'_R \right) - i \epsilon'   \frac{\text{sin}\left(\theta_{k'}/2 \right)}{\eta} F \left( k_R + k'_L \right) \nonumber \\
&+ i \epsilon \; \eta \; \text{sin}\left(\theta_{k'}/2 \right) F \left( - k'_R - k_L \right) + \epsilon \epsilon'  \; \eta \; \text{cos}\left(\theta_{k'}/2 \right) F \left( -k_L + k'_L \right) \bigg] \nonumber \\
\label{eq:bogofermi}
\end{align}
\end{widetext}

where $k_R, k_L, k'_R, k'_L, \omega_R, \omega_L, \Omega_R,$ and $\Omega_L$ are precisely as defined in Eqs.~(\ref{eq:Dopplermomenta}) and~(\ref{eq:Dopplerfrequencies}). 

\subsubsection{Infinite-size limit, chiral excitations populations.}

We now work in the infinite-size limit and analyze the creation of excitations from the vacuum. As before, we look at the creation of massless anti-particles from the massive particle modes sitting in the vacuum (and vice versa) since this conversion precisely amounts to the excitation of the system about the vacuum of the massless modes. In particular, noting that the function $F(x) \rightarrow 2 \pi \delta (x)$ in the limit $L \rightarrow \infty$, we find

\begin{align}
v_{\epsilon', k'} (t' = 0)&=  \text{cos} \left( \frac{\theta_{k'}}{2} \right) \left[ \tilde{u}_{+, k} + \tilde{u}_{-, k} \right]  \bigg|_{k = k^{-1}_R (-k'_R)} \nonumber \\
&- i \text{sin} \left( \frac{\theta_{k'}}{2} \right) \left[ \tilde{u}_{+, k} - \tilde{u}_{-, k} \right] \bigg|_{k = k^{-1}_L (k'_R)} \nonumber \\
& - i \epsilon' \text{sin} \left( \frac{\theta_{k'}}{2} \right) \left[ \tilde{u}_{+, k} + \tilde{u}_{-, k} \right] \bigg|_{k = k^{-1}_R (k'_L)} \nonumber \\
&+ \text{particle content} \propto u_{\pm, k}. 
\label{eq:boundarymatchfermi}
\end{align} 

\begin{center}
\begin{figure}
\includegraphics[width = 3.5in]{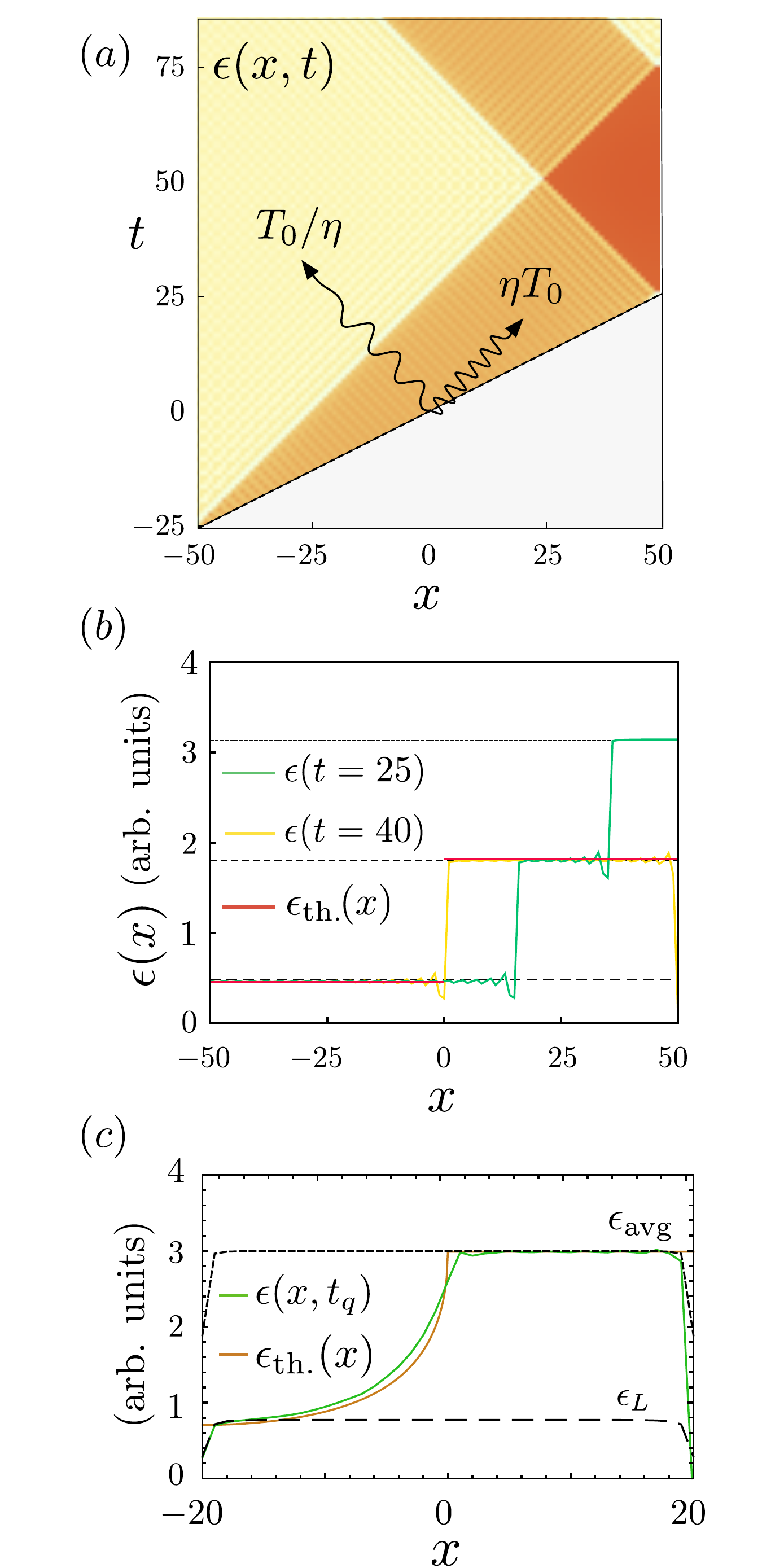}
\caption{(a) The quench is simulated for $v_s = 2$, and the energy density---darker is hotter---is plotted as a function of position and time. The region $x < t$ is seen to be populated with only cold, left-moving waves. The quench ends at time $t_q = 25$. (b) The energy density (arb. units) at two time instants is plotted as a function of position. The dotted lines indicate energy density if all modes are populated according to predicted populations for left-mover/right-movers or an average of the two. (c) he energy density as a function of position at the end of the quench from numerical simulation of the problem in $d = 2$ is compared to the result for $\epsilon_{\text{th.}}$ in the main text.}
\label{fig:heatwavefig}
\end{figure}
\end{center}

The above is a direct result of the integration over the 4 momentum-conserving delta-functions of the Bogoliubov coefficients and one of these terms, $\propto \delta (-k_L -k'_L)$, does not contribute. Two of these terms are associated with the production of right-movers ($\tilde{u}_{R, k} = \left( \tilde{u}_{+,k}  + \tilde{u}_{-,k}\right)/\sqrt{2}$), and one term is associated with left-movers ($\tilde{u}_{L, k} = -i \left( \tilde{u}_{+,k}  - \tilde{u}_{-,k}\right)/\sqrt{2}$). We focus on the first and second terms again since the third term can be shown to be continuously related to the first term, but carries a momentum $k_x > k_0$ while the first carries momentum $k_x < k_0$, where $k_0 = k^{-1}_R ( u_s \gamma_s \sqrt{\abs{\vs{k}_\perp}^2 + m^2})$ is of the order of the mass $m$. 

We now evaluate the energy of the system after the quench. Note that the Hamiltonian $H \equiv \psi^\dagger \cdot h \cdot \psi$, with $h \equiv - i \sigma_z \partial_x$. For $\psi (x, t > x / v_s) = \sum_{\epsilon', k'} \left[ \gamma_{\epsilon', k'} (x,t) f_{\epsilon', k'} + \tilde{\gamma}_{\epsilon', k'} (x,t) \tilde{f}^\dagger_{\epsilon', k'} \right]$, and the state being a vacuum of operators $f_{\epsilon', k'}$ and $\tilde{f}_{\epsilon', k'}$, we find

\begin{center}
\begin{figure}
\includegraphics[width = 3.2in]{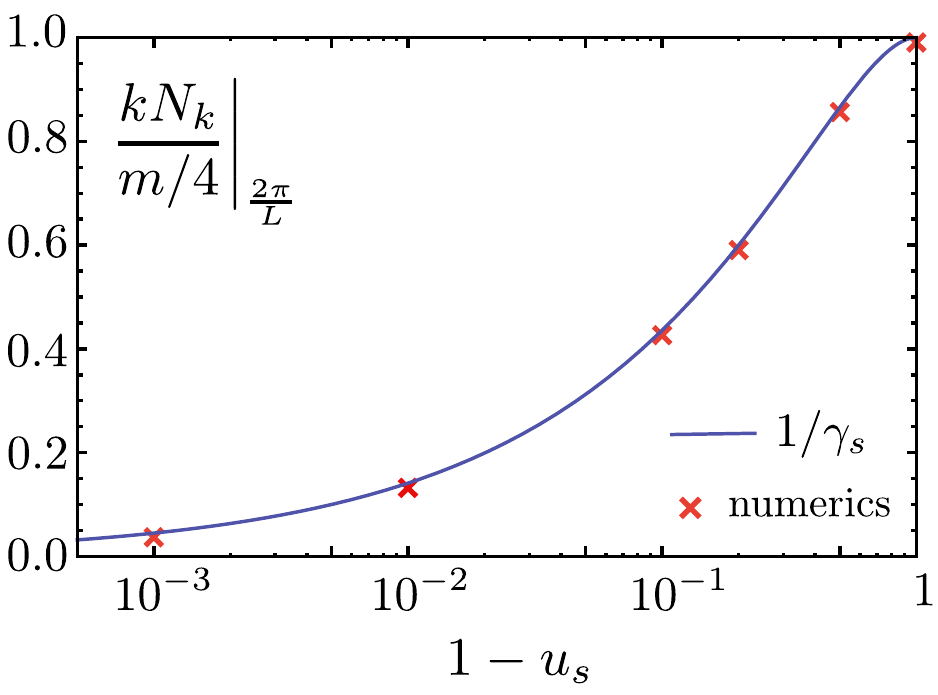}
\caption{Normalized population of the lowest momentum mode, $k = 2 \pi / L$ as opposed to $1/\gamma_s$.}
\label{fig:zeromode}
\end{figure}
\end{center}

\begin{widetext}
\begin{align}
\avg{H} &= \sum \left[ \left( \alpha^{\epsilon_1, \epsilon'}_{k_1, k'} \right)^* \left( \alpha^{\epsilon_2, \epsilon'}_{k_2, k'} \right) \left( \tilde{u}^\dagger_{\epsilon_1, k_1} \cdot h \cdot \tilde{u}_{\epsilon_2, k_2} \right) \right]  + \sum \left[ \left( \beta^{\epsilon_1, \epsilon'}_{k_1, k'} \right) \left( \beta^{\epsilon_2, \epsilon'}_{k_2, k'} \right)^* \left( u^\dagger_{\epsilon_1, k_1} \cdot h \cdot u_{\epsilon_2, k_2} \right) \right] + \sim \tilde{u}^\dagger \cdot h \cdot u + \sim u^\dagger \cdot h \cdot \tilde{u} \nonumber \\
&= \sum \tilde{u}^\dagger_{\epsilon, k} \cdot h \cdot \tilde{u}_{\epsilon, k} - \sum \left( \sum \beta^{\epsilon, \epsilon'}_{k, k'} \tilde{u}_{\epsilon, k} \right)^\dagger \cdot h \cdot \left( \sum \beta^{\epsilon, \epsilon'}_{k, k'} \tilde{u}_{\epsilon, k} \right) + \sum \left( \sum \beta^{\epsilon, \epsilon'}_{k, k'} u_{\epsilon, k} \right)^\dagger \cdot h \cdot \left( \sum \beta^{\epsilon, \epsilon'}_{k, k'} u_{\epsilon, k} \right) \nonumber \\
&\approx \sum_k [ 1 - N^F_L (k) ] (-k) \tilde{u}^\dagger_{L, k} \tilde{u}_{L,k} + \sum_k [ 1 - N^F_R (k) ] (-k) \tilde{u}^\dagger_{R, k} \tilde{u}_{R,k} + \sum_k N^F_L (k) \; k \; u^\dagger_{L, k} u_{L,k} + \sum_k N^F_R (k) \; k \; u^\dagger_{R, k} u_{R,k}. \nonumber \\
\end{align}
\end{widetext}

In the above, all indexes within any brackets are assumed to be summed over. After the first equation, we neglect the oscillating (in time) terms of the form $\tilde{u}^\dagger \cdot u$ and $u^\dagger \cdot \tilde{u}$ which are expected to average out due to the integral over momenta. We also used Eq.~(\ref{eq:bogorelationsfermi}) to simplify the eliminate the $\alpha$ coefficients in favor of the $\beta$ coefficients. The last equation follows by substituting the result of Eq.~(\ref{eq:boundarymatchfermi}) into the second equation while neglecting time-dependent terms of the form $\tilde{u}^\dagger_{\epsilon, k} \tilde{u}_{\epsilon', k'}$ and $u^\dagger_{\epsilon, k} u_{\epsilon', k'}$ which come with $k \neq k'$. The chiral populations are as mentioned in the main text.

\section{Numerical confirmation of heat-wave picture.}

To find the energy-density at any space-time point, we use the picture of ``heat waves" described in Ref.~\cite{Agarwalquantumheatwaves}.The energy density is determined by the average energy carried by waves emanating from the quench front, as illustrated in Fig.~2 (a) of the main text. This picture is verified in the numerical simulations of the above exact results for $d = 1$ and $d =2$, as presented in Fig.~(\ref{fig:heatwavefig}). 

\section{Infinite accuracy in state preparation.}

We now examine the population of the lowest momentum modes (with momentum $k \sim 2 \pi / L$) in the supersonic quench protocol and show that this goes to zero as $\gamma_s \rightarrow \infty$ even for the finite system. This shows that our protocol can be used to produce a state that is arbitrarily close in energy to the ground state. First, we note that the analysis in Sec.~\label{sec:infinitsize} assumes the infinite size limit before discussing the consequences of the limit $v_s \rightarrow 1^+$; there we can discuss the populations of left-moving and right-moving modes separately. For finite system size, and at the lowest momenta, there is no distinction between the population left- and right-moving modes. Instead, we will directly calculate the excited population $N_k$ via the relation $N_k = \sum_{k', \epsilon'} \abs{\beta^{\epsilon, \epsilon'}_{k,k'}}^2$. 

To evaluate this sum we first note that one may again decompose $\beta_{k,k'}$ in terms of a sum of delta functions where the continuum limit is reached by taking $\gamma_s \rightarrow \infty$ instead of $L \rightarrow \infty$. Consider the modification of $F(x) \rightarrow F(x' \gamma^2_s)$, with $x' = x/\gamma^2_s$. First, the parameter $x'$ is, for instance, $k'_R / \gamma^2_s$ and $k'_L / \gamma^2_s$. Then, $\abs{dx' / dn} = \frac{1}{\gamma_s} \left( 1 \pm \frac{k'}{\Omega_{k'}} \right) \times \frac{2 \pi}{L}$, where $n$ is the discrete index enumerating the modes. As $\gamma_s \rightarrow \infty$, $x'$ clearly becomes continuous. Moreover, $F(x) \sim \sin (x' \gamma_s L / 2)$; as $\gamma_s \rightarrow \infty$, this becomes a delta-function due to the squeezing of the sinc function. The above discussion establishes that the limit $\gamma_s \rightarrow \infty$ acts in the same spirit as $L \rightarrow \infty$. It remains to evaluate the population of the modes by integrating over the delta functions in $\sum_{k', \epsilon'} \abs{\beta^{\epsilon, \epsilon'}_{k,k'}}^2$. Using the result $\int \delta^2 ( x - x') f(x') dx' = f(x) \delta (0)$, where in this case, $\delta (0) \equiv (\gamma_s L)/(2 \pi)$, we find 

\be
N_{\epsilon = \pm, k \sim 1/L} = \frac{m}{4 \gamma_s}
\label{eq:lowmomentum}
\ee

Thus, the population of the lowest momentum modes goes to zero as $v_s \rightarrow 1^+$. Thus, one can reach a state arbitrarily close to the ground state in the Lorentz cooling protocol. Moreover, since protocol takes time $L/v_s$ plus a few multiples of the quench time-scale $\tau$ which we assume to be $\mathcal{O} (L^0)$, it is more efficient in time than the adiabatic protocol. We numerically verify Eq.~(\ref{eq:lowmomentum}) for $L = 100, m = 2 \pi$ in Fig.~\ref{fig:zeromode}. 


%